\documentclass[11pt,a4paper]{article}
\usepackage{jcappub}
\usepackage{amsmath}
\usepackage{subcaption}
\usepackage{graphicx}
\usepackage{tikz}
\usepackage{xcolor}

\makeatletter
\gdef\@fpheader{}
\makeatother

\newlength{\fullw}
\newlength{\halfw}
\newlength{\threefigw}
\newlength{\twofigw}
\newlength{\onefigw}
\newcommand{\order}[1]{\mathcal{O}\!\left(#1\right)}
\newcommand{\heaviside}[1]{\mathrm{\Theta}\!\left( #1 \right)}
\newcommand{\heavisideb}[1]{\mathrm{\Theta}\!\left[ #1 \right]}

\newcommand{\horizon}[1]{\ud_{\mathrm{h}}\negthinspace\left(#1\right)}

\newcommand{\ud}{\mathrm{d}}

\newcommand{\uc}{\mathrm{c}}
\newcommand{\uv}{\mathrm{v}}
\newcommand{\up}{\mathrm{p}}
\newcommand{\uini}{\mathrm{ini}}
\newcommand{\ucur}{\mathrm{cur}}
\newcommand{\ucmb}{\mathrm{cmb}}
\newcommand{\umat}{\mathrm{mat}}
\newcommand{\urad}{\mathrm{rad}}
\newcommand{\ucorr}{\mathrm{corr}}
\newcommand{\ueq}{\mathrm{eq}}
\newcommand{\utot}{\mathrm{tot}}
\newcommand{\uK}{\mathrm{K}}
\newcommand{\uprod}{\mathrm{prod}}
\newcommand{\urel}{\mathrm{rel}}

\newcommand{\calJ}{\mathcal{J}}

\newcommand{\calN}{\mathcal{N}}
\newcommand{\calP}{\mathcal{P}}

\newcommand{\calR}{\mathcal{R}}
\newcommand{\calT}{\mathcal{T}}

\newcommand{\Jpsd}{\calJ_\textsc{psd}}

\newcommand{\TeV}{\mathrm{TeV}}

\newcommand{\tstar}{t_\star}
\newcommand{\tini}{t_\uini}
\newcommand{\tcur}{t_\ucur}
\newcommand{\teq}{t_\ueq}
\newcommand{\tzero}{t_0}
\newcommand{\tell}{t_\ell}
\newcommand{\ty}{t_{\lambda y}}
\newcommand{\Hzero}{H_0}

\newcommand{\zcur}{z_\ucur}
\newcommand{\zini}{z_\uini}

\newcommand{\Tcur}{T_\ucur}
\newcommand{\Tcmb}{T_\ucmb}
\newcommand{\Tini}{T_\uini}

\newcommand{\q}{q}
\newcommand{\qcur}{\q_\ucur}
\newcommand{\qzero}{\q_0}
\newcommand{\qini}{\q_\uini}

\newcommand{\N}{\dfrac{\ud \calN}{\ud \ell}}
\newcommand{\NN}{\dfrac{\ud \calN}{\ud N}}
\newcommand{\Nn}{\dfrac{\ud^2 \calN}{\ud \ell \ud N}}

\newcommand{\Cini}{C_\uini}

\newcommand{\rv}{r_0}
\newcommand{\vort}{\mathrm{vort}}
\newcommand{\rel}{\mathrm{vort, \urel}}
\newcommand{\pro}{\mathrm{vort, \uprod}}
\newcommand{\doom}{\mathrm{doom}}
\newcommand{\proto}{\mathrm{proto}}

\newcommand{\xmax}{x_{\max}}
\newcommand{\xmin}{x_{\min}}
\newcommand{\ymax}{y_{\max}}
\newcommand{\ymin}{y_{\min}}
\newcommand{\yeq}{y_\ueq}

\newcommand{\ellT}{\ell_\textsc{t}}
\newcommand{\ellbarT}{\bar{\ell}_\textsc{t}}
\newcommand{\ellbareq}{\bar{\ell}_\ueq}
\newcommand{\ellini}{\ell_\uini}
\newcommand{\ellbarcur}{\bar{\ell}_\ucur}
\newcommand{\ellbarini}{\bar{\ell}_\uini}
\newcommand{\ellcorr}{\ell_\ucorr}

\newcommand{\lcur}{\ell_\ucur}
\newcommand{\lcorr}{\ellcorr}

\newcommand{\tsigma}{\tcur}
\newcommand{\U}{\mu}
\newcommand{\T}{\calT}
\newcommand{\NNstar}{\calR}

\newcommand{\gammad}{\Gamma G \U}
\newcommand{\gammac}{\gamma_\uc}
\newcommand{\gammav}{\gamma_\uv}

\newcommand{\nurad}{\nu_\urad}
\newcommand{\numat}{\nu_\umat}

\newcommand{\Mp}{M_\mathrm{Pl}}
\newcommand{\OmegaDM}{\Omega_{\textsc{dm}}}

\newcommand{\Omegatot}{\Omega_{\utot}}
\newcommand{\Omegaprod}{\Omega_{\uprod}}
\newcommand{\Omegarel}{\Omega_{\urel}}
\newcommand{\Omegarelmin}{\Omegarel^{\min}}

\newcommand{\ellv}{\ell_\uv}
\newcommand{\ellp}{\ell_\up}

\definecolor{applegreen}{rgb}{0.0,0.5,0.0}
\definecolor{imperialred}{rgb}{0.93,0.16,0.22}

\author[a]{Pierre Auclair,}
\author[b,c]{Patrick Peter,}
\author[d,b]{Christophe Ringeval}
\author[a]{and Dani\`ele Steer}

\affiliation[a]{Laboratoire Astroparticule et Cosmologie, Universit\'e de Paris, 10 rue Alice Domon et L\'eonie Duquet, 75013
  Paris, France}
\affiliation[b]{${\cal G}\mathbb{R}\varepsilon\mathbb{C}{\cal O}$ -- Institut
d'Astrophysique de Paris, CNRS \& Sorbonne Universit\'e, UMR 7095
98 bis boulevard Arago, 75014 Paris, France}
\affiliation[c]{Centre for Theoretical Cosmology, Department of Applied
Mathematics and Theoretical Physics, University of Cambridge, Wilberforce
Road, Cambridge CB3 0WA, United Kingdom}
\affiliation[d]{Cosmology, Universe and Relativity at Louvain,
  Institute of Mathematics and Physics, Louvain University, 2 Chemin
  du Cyclotron, 1348 Louvain-la-Neuve, Belgium}
\emailAdd{auclair@apc.in2p3.fr}
\emailAdd{peter@iap.fr}
\emailAdd{christophe.ringeval@uclouvain.be}
\emailAdd{steer@apc.in2p3.fr}

\title{Irreducible cosmic production of relic vortons}

\begin{document}

\abstract{The existence of a scaling network of current-carrying
  cosmic strings in our Universe is expected to continuously create
  loops endowed with a conserved current during the cosmological
  expansion. These loops radiate gravitational waves and may stabilise
  into centrifugally supported configurations. We show that this
  process generates an irreducible population of vortons which has not
  been considered so far. In particular, we expect vortons to be
  massively present today even if no loops are created at the time of
  string formation. We determine their cosmological distribution, and
  estimate their relic abundance today as a function of both the
  string tension and the current energy scale. This allows us to rule
  out new domains of this parameter space. At the same time, given
  some conditions on the string current, vortons are shown to provide a
  viable and original dark matter candidate, possibly for all values
  of the string tension. Their mass, spin and charge spectrum being
  broad, vortons would have an unusual phenomenology in dark
  matter searches.}


\maketitle

\section{Introduction}

Cosmic strings are expected to be formed in most extensions of the
standard particle physics model as stable line-like topological
defects formed during high temperature, $\Tini$ say, symmetry breaking
phase transitions in the early Universe~\cite{Kibble:1976sj}. This
occurs whenever a symmetry $\mathcal{G}$ is broken down to a smaller
one $\mathcal{H}$ provided the first homotopy group of the quotient
group $\mathcal{G}/\mathcal{H}$ (vacuum manifold) is non-trivial,
producing similarly non-trivial topological solutions for the
symmetry-breaking Higgs field. The scaling evolution of cosmic string
networks (see e.g. Ref.~\cite{Lorenz:2010sm} and references therein)
means that they are present throughout the evolution of the Universe,
possibly giving rise to numerous different observational signatures,
such as line-like discontinuities in temperature in the Cosmic
Microwave Background (CMB), or bursts of gravitational
waves~\cite{Damour:2001bk,Blanco-Pillado:2017oxo,Ringeval:2017eww}. These
very much sought-for signatures in turn lead to strong constraints on
the string tension $G\mu$.

Most studies of cosmic strings suppose they are structureless, with
equal energy per unit length and tension, and therefore they are
expected to be well described by a no-scale 2-dimensional worldsheet
action, i.e. the Nambu-Goto action.  This is no longer the case if, as
first realised by Witten \cite{Witten:1985fp,Lazarides:1986di},
particles coupled to the string-forming Higgs field can condense in the
string core and subsequently propagate along the worldsheet.
The resulting strings thus behave like current carrying wires and are
endowed with a much richer structure~\cite{Carter:1999pq,Carter:2000wv}.

One of the simplest examples of current-carrying strings is that of a
U$(1)_R \times $U$(1)_Q$ gauge theory with an unbroken gauge symmetry
$Q$ (which might be electromagnetism, but not necessarily) and a
broken symmetry $R$ \cite{Witten:1985fp}. This model generalises the
proto-typical Abelian-Higgs model of cosmic strings behind much of the
existing work on cosmic strings. At a temperature $\Tini$, and a
cosmic time $\tini$, the Higgs field $\phi$ with $Q=0$ and $R=1$
acquires a non-zero vacuum expectation value $|\langle\phi\rangle|
\neq 0$, thereby breaking the first component U$(1)_R$ of the total
invariance group; this leads to the formation of vortex lines. The
field $\phi$ vanishes at the core of the string and its phase varies
by an integer times $2\pi$ along any closed path around the vortex:
this is the standard Kibble mechanism. If the theory contains fermions
obtaining their masses from the U$(1)_Q$ broken symmetry, those
form zero modes in the string core where the symmetry is restored, 
thereby forming a superconducting current.

The model also comprises a
second scalar field $\sigma$ with $Q=1$ and $R=0$, the coupling
potential between $\phi$ and $\sigma$ being chosen such that $\langle
\sigma\rangle = 0$ in vacuum (where $|\langle\phi\rangle| \neq
0$). Under certain conditions, it is energetically favourable to have
$\langle \sigma \rangle \neq 0$ at the core of the string where
$\langle\phi\rangle = 0$.  At a temperature $\Tcur < \Tini$, and
cosmic time $\tcur > \tini$, the charged scalar field $\sigma$ thus
condenses on the string and acts as a bosonic charge carrier making
the string current-carrying (and in fact actually superconducting).
In the present paper, we assume that the current sets in long after
the string formation scale. In the language of
Refs.~\cite{Kibble:1981gv, Brandenberger:1996zp}, this means we assume
the current is formed long after the friction damping regime has
finished, i.e. during the radiation era. In practice, it means that we
consider $\Tini$ (and $\tini$) to be the end of the friction dominated
regime.

Cosmic strings can also be produced~\cite{Jones:2002cv,Sarangi:2002yt}
in superstring theory, also forming, under specific conditions, a
network similar to a Nambu-Goto network~\cite{Urrestilla:2007yw}.
Whether or not these so-called cosmic superstrings can carry a current
deserves more investigation since they have been shown to not be able
to hold fermionic zero modes so that only bosonic condensates can
source such a current~\cite{Polchinski:2004yav}.  It should, however,
be mentioned that because cosmic superstrings live in a higher
dimensional manifold, their motion in the extra dimensions projected
into the ordinary 3 dimensional space should be describable by means
of a phenomenological non-trivial equation of
state~\cite{Carter:1990bb,Carter:1994hn} mimicking that of a
current-carrying string; this can be interpreted as moduli field
condensates.

The presence of currents flowing along the strings affects the
dynamics of the network, and in this paper we particularly focus on
vortons~\cite{Davis:1988ij,Carter:1990gz,Brandenberger:1996zp,Martins:1998gb,Martins:1998th,Carter:1999wy,Davis:2000cx,Steer:2000jn},
namely closed loops of string which are stabilised by the angular
momentum carried by the current. Vortons do not radiate classically,
and here we make the assumption that they are classically stable
as well (see for instance \cite{Lemperiere:2003yt, Battye:2008mm,
Garaud:2013iba} for numerical studies of their stability). On cosmological scales, they appear as point particles
having different quantized charges and angular momenta.

In this work, we extend the derivation of the vorton abundance of
Ref.~\cite{Brandenberger:1996zp} by not only considering vortons
produced from pre-existing loops \emph{at} $\tini$, but also those
vortons that may form from the loops chopped off the network at all
subsequent times. In particular, we extend the work of
Ref.~\cite{Peter:2013jj}, in which a Boltzmann equation governing the
vorton density has been derived and integrated for any loop production
function (LPF), but not explicitly solved to get cosmological
constraints. Let us notice that some of these new produced vortons,
when created from the network, may be highly boosted. However,
extrapolating the mean equation of state obtained for Nambu-Goto
cosmic string loops, their momentum gets redshifted away and, on
average, they behave as non-relativistic
matter~\cite{Ringeval:2005kr}. For this reason, the produced vortons
are, as those originally considered in
Ref.~\cite{Brandenberger:1996zp}, potential dark matter and cosmic
rays candidates~\cite{Carter:1990gz, Bonazzola:1997tk}.

The total abundance of vortons today is expected to depend on $\tcur$
as well as $\tini$, and hence on the underlying particle physics
model. Determining their density parameter today, say $\Omegatot$, and
using the current constraints on $\OmegaDM h^2 \simeq 0.12$ will allow
us to place constraints on the physics at work in the early
Universe~\cite{Aghanim:2018eyx}.

The formation and build-up of a population of vortons can be studied
using a Boltzmann equation~\cite{Peter:2013jj}.  In this paper, we
extend this work by applying the framework introduced
in~\cite{Auclair:2019jip} to estimate quantitatively the density of
vortons today. In section \ref{sec:assumptions} below, we review the
necessary physics underlying vorton properties, then in
section~\ref{sec:dist}, we evaluate the distribution of loops and
vortons, in order to be able to calculate, in section~\ref{sec:vorts},
the actual vorton distribution and, finally, their relic abundance in
section~\ref{sec:DM}. We end this work by some concluding remarks.

\section{Assumptions on the physics of vortons}\label{sec:assumptions}

As discussed in the introduction, we focus in this paper on cosmic
strings that emerged at a temperature $\Tini$ and later became current
carrying at a temperature $\Tcur$.

For non-conducting strings, the boost invariance along the string
implies that the string tension $\T$ and its energy per unit
length $\mu$ are equal and, in order of magnitude, given by
$\U=\T = m_\phi^2$, where $m_\phi \propto |\langle \phi
\rangle|$ is the mass of the string-forming Higgs field $\phi$.  As
soon as a current flows along the string, the worldsheet Lorentz
invariance is broken and so is the degeneracy between the
stress-energy tensor eigenvalues $\mu$ and
$\T$~\cite{Carter:1989dp,Carter:1992ny,Carter:2000wv}, the
tension being reduced and the energy per unit length increased by the
current in such a way that
\begin{equation}
\T < m_\phi^2 < \U.
\end{equation}
The equation of state of current-carrying strings \cite{Babul:1987me,Peter:1992dw,Carter:1994hn,Carter:1999hx,Ringeval:2000kz,Ringeval:2001xd} provides us with a saturation condition
\begin{equation}
    \U - \T \leq m_\sigma^2 \implies 0 < \frac{\U - \T}{m_\phi^2}
    \leq \dfrac{m_\sigma^2}{m_\phi^2}\,,
    \label{UTm}
\end{equation}
according to which there exists a maximal spacelike current, above
which it becomes energetically favoured for the condensate to flow out
of the string.  For a timelike
current~\cite{Peter:1992dw,Peter:1992nz}, i.e.~a charge, there exists
a phase frequency threshold allowing, in principle, for arbitrary
large values of the charge.  However, vacuum polarisation effectively
reduces the integrated charge \cite{Peter:1992ta} so that saturation
holds for all possible situations.

Denoting by $\lambda$ the Compton wavelength of the current carrier
($\lambda \simeq m_\sigma^{-1}$), we define the parameter $\calR$ by
\begin{equation}
\calR \equiv \lambda \sqrt{\U}\,.
\label{calR}
\end{equation}
Because $\U \simeq m_\phi^2$, this quantity is approximately the ratio
between the Compton wavelengths of the current carrier and the one of
the string forming Higgs field, or, equivalently, $\calR \simeq
m_\phi/m_\sigma$ which we assume to be greater than unity. Given
\eqref{UTm}, it is safe to assume that, at least for $\calR \gg 1$,
the string tension and the energy per unit length are numerically so
similar that distinguishing between them is irrelevant in the
forthcoming cosmological context; we will thus denote them both by the notation $\U$.

A current-carrying closed string loop is characterized by two
classically conserved integral quantum numbers $N$ and $Z$, generally
non-zero, which prevent the loop from disappearing
completely~\cite{Carter:1990sm}.  As the loop loses energy through
friction or radiation, it reaches a classically stable state called a
\emph{vorton}~\cite{Davis:1988ij}.  However, this state can decay
through quantum tunnelling if the size of the loop is comparable with
the Compton wavelength of the current carrier, $\lambda$. Hence a
vorton can only be stable if the current flowing along the string loop
can prevent its collapse and if its proper length is much larger than
$\lambda$.

Although the values of $N$ and $Z$ are initially randomly distributed,
it is expected that the majority of closed loops are of nearly
\emph{chiral}~\cite{Davis:1997bs,Martins:1998th,Carter:1999hx,Carter:2003fb}
type with almost identical quantum numbers~\cite{Carter:1990sm}.
Besides, the loop rotation velocity $v_\vort = \sqrt{\T/\U}\simeq 1$
is roughly approximated by that of light and
\begin{equation}
    |Z| \approx N.
\end{equation}
In the rest of the paper, we focus on such nearly chiral vortons.
Using of the central limit theorem, we estimate that the value of $N$ at the formation of a loop is given by
\begin{equation}
    N_\star = \sqrt{\frac{\ell_\star}{\lambda}}.
    \label{Nstar}
\end{equation}
In \eqref{Nstar} and in the rest of this paper, a subscript $\star$ on a quantity denotes the value it had at the time of formation of the corresponding loop.
Since the charge $N$ is conserved, we can, in what follows, omit the index $\star$ and simply write $N_\star = N$.

To estimate the size of the vortons $\ell_0$, we first have to note
that they have been shown to approach circularity
\cite{Lemperiere:2003yt}.  Moreover, large vortons would also tend to
circularize through either gravitational or gauge field radiation, on time
scales much smaller than the Hubble time.  It thus seems reasonable to
consider mostly circular loops, therefore described by one parameter
only, namely their radius $\rv = \ell_0 / 2\pi$.  Vortons are also
characterised by their angular momentum quantum number $J = NZ \approx
N^2$.  Equivalently, it is also given in terms of the energy per unit
length and tension by~\cite{Carter:1990gz} $J = 2\pi \rv^2
\sqrt{\T\mu}$, i.e. $J^2 = \mu \T \ell_0^4 / (4\pi^2)$.  Hence for
chiral vortons with $\calR \gg 1$
\begin{equation}
    \ell_0 = \sqrt{\frac{2\pi}{\mu}} N= \sqrt{\frac{2 \pi \ell_\star}{\lambda \mu}} \approx \sqrt{\frac{\ell_\star}{\lambda \mu}},
\end{equation}
provided $\ell_0 > \lambda$. The length $\ell_0(N)$ being itself a
function of the charge $N$, this is equivalent to imposing that $N >
\calR$. Therefore, $\calR$ gives also the minimal possible charge of a
vorton.

Following the same procedure as in \cite{Auclair:2019jip}, we model
the physics of the vortons using an arbitrary function $\calJ$ which
describes how the current-carrying loops lose energy
\begin{align}
    \frac{\ud \ell}{\ud t} &= - \Gamma G\mu \calJ(\ell,N),
    \label{eq:ldot}
    \\
    \frac{\ud N}{\ud t} &= 0,
\end{align}
in which $\Gamma \approx 50$ is a numerical factor for the emission of
gravitational waves (GW)~\cite{PhysRevD.45.1898}. In order to model
string networks with vortons, we impose the following properties on $\calJ$:
\begin{itemize}
    \item $\calJ(\ell \gg \ell_0, N) \approx 1$, meaning that on scales much larger than the vorton size, the effect of the current is mostly negligible so that the dynamics of the current-carrying string is well approximated by that of a Nambu-Goto string; gravitational wave radiation is the dominant energy-loss mechanism and we neglect other such mechanisms.
    \item $\calJ(\ell \ll \ell_0, N) \approx 0$ if $\ell_0 > \lambda$, meaning that the angular momentum carried by the current prevents the loop from shrinking, provided the loop is large enough to prevent quantum tunnelling.
\end{itemize}
We will consider a smooth form of $\calJ$, regulated by a parameter $\sigma$, in particular
\begin{align}
    \calJ(\ell,N) = \frac{1}{2}\left\{1 + \tanh\left[\frac{\ell - \ell_0(N)}{\sigma}\right]\right\}.
\end{align}
We call \emph{vortons} all the loops with sizes $\ell \leq \ell_0(N)$ and $N > \calR$.
In the limit $\sigma \rightarrow 0$,  $\calJ(\ell,N)$ reduces to $\Theta[\ell - \ell_0(N)]$, and the vortons accumulate around $\ell_0(N)$.

Let us mention that our approach, and results, differ from the vorton
abundances derived in Refs.~\cite{Martins:1998th,
  Martins:1998gb}. These latter references were concerned with the
extreme limit in which the current carrier condensation and string
forming times are similar ($\calR \simeq 1$ in our notation). For this
reason, they were not concerned with the emission of gravitational
waves. Indeed, in the limit $\calR\to 1$, strong currents have been
shown to dampen the loop oscillations and this allows for a population
of vortons to be rapidly created (soon after the string forming phase
transition). The vortons considered in Refs.~\cite{Martins:1998th,
  Martins:1998gb} are of this kind only. Let us recall that the
current-carrier particles are trapped on the string worldsheet by
means of a binding potential. As such, when there are strong currents,
there is always the possibility that they tunnel
out~\cite{Peter:1992dw}. Such an instability could drastically affect
the current, and hence the mechanism by which the vortons considered
in Refs.~\cite{Martins:1998th, Martins:1998gb} are formed. On the
contrary, the vortons we are considering here carry weak currents and
our results are only valid in the domains for which $\calR > 1$. The
damping mechanism by which the weak current-carrying loops become
vortons is the emission of gravitational waves (as in
Ref.~\cite{Brandenberger:1996zp}).

Having recalled the basic properties of vortons and their dynamics, we
now turn to the expected distributions of loops of various kinds,
including those ending up as vortons.

\section{Distribution of loops and vortons}\label{sec:dist}

\begin{figure}
\begin{center}
    \begin{tikzpicture}
        \draw [very thick] [->] (0,0) -- (12,0);
        \draw (1, 0) node [below] {$\tini$};
        \draw [dotted, thick] (1,-0.2) -- (1,0.2);
        \draw (3.5, 0) node [above] {standard NG strings};
        \draw (6, 0) node [below] {$\tcur$};
        \draw [dotted, thick] (6,-0.2) -- (6,0.2);
        \draw (9, 0) node [above] {current carrying strings};
        \draw (12, 0) node [right] {$t$};
    \end{tikzpicture}
\end{center}
    \caption{At time $\tini$ and temperature $\Tini$ a network of
      strings forms with an initial distribution. At the later time
      $\tcur$ the strings become current-carrying, and vortons can
      form. At all times, loop can be produced from long strings and
      larger loops with a given loop production function.}
    \label{fig:dani-notrubbish}
\end{figure}
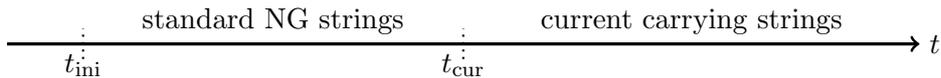

In the following sections, we extend a statistical method
originally based on the Boltzmann equation~\cite{Copeland:1998na,
Lorenz:2010sm, Rocha:2007ni, Peter:2013jj} to study current carrying strings. Our aim is to find the
number density of vortons, marginalized over their charge $N$, with
length $\ell$ at time $t>\tcur$, given some initial loop distribution
at time $\tini$ and some assumptions about the loop production
function (see figure~\ref{fig:dani-notrubbish}).

\subsection{Continuity equation for the flow of loops in phase space}

Let $\ud^2\calN(\ell,t,N)/\ud\ell \ud N$ be the number density of
loops with length $\ell$ and charge $N$ at time $t$.  In an expanding
universe with scale factor $a(t)$, and taking into account the fact
that loops lose length at a rate which depends on their length as
expressed through equation~\eqref{eq:ldot}, the continuity equation
for the number density of loops is given by
\cite{Copeland:1998na,Auclair:2019jip}
\begin{equation}
    \frac{\partial}{\partial t} \left[ a^3 \Nn(\ell,t,N) \right] - \Gamma G\mu \frac{\partial}{\partial \ell} \left[ a^3 \calJ(\ell,N) \Nn(\ell,t,N)\right] = a^3\calP(\ell,t,N).
    \label{eq:main}
\end{equation}
Here $\calP(\ell,t,N)$ is the charged loop production function (LPF),
namely the rate at which loops of length $\ell$ and charge $N$ are
formed at time $t$ by being chopped off the string network and we will
specify it below. Note that this equation is exactly equivalent to
that of Ref.~\cite{Peter:2013jj}, as we explain in details in
Appendix~\ref{App1}.

The solution to equation~(\ref{eq:main}) can be obtained in integral form
following a similar procedure to that explained in
Ref.~\cite{Auclair:2019jip}, though one must take into account the new
independent variable $N$. Upon multiplying by $\calJ(\ell,N)$,
equation~\eqref{eq:main} becomes
\begin{equation}
    \frac{\partial g}{\partial t}(\ell,t,N) - \Gamma G\mu \calJ(\ell,N) \frac{\partial g}{\partial \ell} (\ell,t,N) = a^3(t) \calJ(\ell,N) \calP(\ell,t,N),
    \label{eq:c}
\end{equation}
where we have defined
\begin{equation}
    g(\ell,t,N) \equiv a^3 \calJ(\ell,N) \Nn \,.
    \label{eq:gdef}
\end{equation}
The change of variables $\left\{\ell,t,N\right\} \to \left\{\xi,\tau,N\right\}$, with
\begin{equation}
   \xi \equiv \int \frac{\ud \ell}{\calJ(\ell,N)} \qquad \hbox{and} \qquad \tau \equiv \Gamma G \mu t,
\end{equation}
enables equation~(\ref{eq:c}) to be written in the simpler form
\begin{equation}
    \frac{\partial g(\xi,\tau,N) }{\partial \tau}- \frac{\partial g(\xi,\tau,N)}{\partial \xi} = \frac{a^3(\tau)}{\Gamma G\mu} \calJ(\xi,N) \calP(\xi,\tau,N).
\end{equation}
Upon using \emph{light cone} type coordinates
\begin{equation}
     u \equiv \frac12 \left( \tau - \xi\right) \qquad \hbox{and} \qquad
    v \equiv \frac12 \left(\tau + \xi\right),
    \label{eq:uv}
\end{equation}
it follows that equation~\eqref{eq:main} reduces to
\begin{equation}
    \frac{\partial g(u,v,N) }{\partial u}= \frac{a^3(u,v)}{\Gamma G\mu} \calJ(u,v,N) \calP(u,v,N),
    \label{eq:here}
\end{equation}
which can be integrated between $\tcur$ and $t$, or in terms of the
variable $u = -v+\tau = -v + \Gamma G\mu t$, between $u_\ucur =
-v + \tau_\ucur = -v + \Gamma G\mu \tcur$ to $u$,
\begin{equation}
    g(u,v,N) - g(-v + \Gamma G\mu \tcur,v,N) = \int_{-v + \Gamma G\mu \tcur}^u \! \frac{a^3(u',v)}{\Gamma G\mu} \calJ(u',v,N) \calP(u',v,N)\,\ud u',
    \label{eq:relaxsol}
\end{equation}
the integral in equation~\eqref{eq:relaxsol} being calculated with $v$
constant. Rewritten in terms of $\ud^2\calN(\ell,t,N)/\ud\ell \ud N$ using
equation~\eqref{eq:gdef} finally gives
\begin{equation}
\begin{aligned}
    a^3(t) \calJ(\ell,N) \Nn & =  a^3(\tcur)
    \calJ\left(\lcur,N\right) \Nn\left(\lcur,\tcur,N\right)  \\
    & + \int_{-v + \Gamma G\mu \tcur}^u \! \frac{a^3(u',v)}{\Gamma G\mu} \calJ(u',v,N) \calP(u',v,N) \, \ud u'.
    \label{eq:sol1}
\end{aligned}
\end{equation}
Here $\lcur$ is the size of the loops at condensation and is a
function $\lcur(\ell,t,N)$. It is found using the variable $v = \tau +
\xi$ of equation~\eqref{eq:uv} which is a constant along the flow,
namely $\lcur$ is a solution of
\begin{equation}
    \xi(\lcur,N) = \xi(\ell,N) + \Gamma G\mu (t-\tcur).
    \label{eq:lcur}
\end{equation}

The solution of the continuity equation~\eqref{eq:main} is
therefore given by equation~\eqref{eq:sol1}. On the right-hand-side,
we recognise two terms. The first are the loops left over from the
pre-existing loop distribution at the time of condensation,
$t=\tcur$. The second term contains those loops which are produced
from the string network at time $t>\tcur$.
As we will see in more detail in section~\ref{sec:vorts}, each of
these distributions contain three kinds of
loops~\cite{Brandenberger:1996zp}:

\begin{enumerate}
    \item \emph{Doomed loops}: these loops have an initial size which
      is too small to support a current, and hence they decay through
      gravitational radiation never becoming vortons. They are
      characterised by quantum numbers $N<\calR$.
    \item \emph{Proto-vortons}: these are loops which are initially
      large enough to be stabilised by a current (thus $N>\calR$), but
      have not yet reached the vorton size $\ell_0$.
    \item \emph{Vortons}: these are all those proto-vortons which have
      decayed by gravitational radiation to become vortons. Hence
      vortons have $N>\calR$, and in the limit $\sigma \rightarrow 0$,
      they accumulate with length $\ell_0(N)$.
\end{enumerate}

Our aim in the following is to extract these different
distributions. Each will contain two contributions: those formed from
the initial distribution i.e.~coming from the first term in
equation~\eqref{eq:sol1}, and those produced at later times from
being chopped off the string network, i.e.~coming from the
second term in equation~\eqref{eq:sol1}. In the case of vortons, we
call these two families ``relaxed vortons'' and ``produced vortons'',
respectively. In section~\ref{sec:DM}, we will use these to determine
their relic density and put constraints on $G\mu$ and $\calR$.

\subsection{The loop distribution at condensation}
\label{sec:icloop}

A first step is to specify the loop distribution at $\tcur$. The
strings are assumed to form at a temperature $\Tini$ corresponding to
a time $\tini$ in the early Universe. At all times $\tini<t<\tcur$,
that is before condensation, they behave as standard Nambu-Goto
strings, see figure~\ref{fig:dani-notrubbish}. Hence the loop
distribution is the canonical one, i.e.~contains a population of loops
formed at $\tini$ and another population of scaling loops created from
the long strings and larger loops~\cite{Auclair:2019zoz}.

The main simplifying assumption of our work is to assume a Dirac
distribution for the loop production function, namely
\begin{equation}
  \calP(\ell,t)=C t^{-5} \delta\negthinspace\left(\frac{\ell}{t}
  - \alpha \right),
  \label{eq:LPF-before}
\end{equation}
with $C=1$ and $\alpha = 0.1$ as to match the Kibble, or one scale,
model~\cite{Kibble:1976sj}. Hence all the produced loops that are chopped
off the network are assumed to be of the same size, given by the
fraction $\alpha$ of $t$, which is, up to a constant of order unity,
the horizon size. This assumption allows us to analytically
solve for the produced vorton distribution later on. However, we
stress that more realistic loop production functions, such as the
Polchinski-Rocha one~\cite{Polchinski:2006ee, Dubath:2007mf,
  Rocha:2007ni, Lorenz:2010sm, Auclair:2020oww}, produce smaller loops
while matching in amplitude with the Dirac LPF for
$\ell/t=\alpha$~\cite{Ringeval:2005kr,Auclair:2019zoz}. Therefore, when gravitational
wave emission from loops is accounted for (which is the case here),
the resulting scaling loop distributions end up being quite similar over the length scales $\ell > \gammad t$. They may, however, differ
significantly on smaller length scales, namely for $\gammac t <\ell <
\gammad t$, where $\gammac$ stands for the length scale at which
gravitational backreaction damps the LPF~\cite{, Lorenz:2010sm}. For
Nambu-Goto strings, this length scale is expected to verify $\gammac
\ll \gammad$~\cite{Siemens:2002dj, Polchinski:2007rg}. Therefore, our
results derived here from a Dirac LPF should provide a robust lower
bound for all the others LPF, and may also be directly applicable to
the Polchinski-Rocha ones but only in the limit in which $\gammac \simeq
\gammad$.

Under these assumptions, the resulting distribution of cosmic string
loops at time $\tcur$ is given by~\cite{Auclair:2019zoz}
\begin{equation}
\begin{aligned}
    \N\left(\ell,\tcur\right) & = C \, \tcur^{-3/2} \frac{ (\alpha +
      \Gamma G\mu)^{3/2}}{(\ell + \Gamma G\mu \tcur)^{5/2}}
    \Theta(\alpha \tcur-\ell)\Theta\left[\ell + \Gamma G\mu
      \tcur-\tini(\alpha + \Gamma G\mu)\right] \\ & + \Cini
    \left(\frac{\tini}{\ell}\right)^{5 / 2} \tini^{-4}
    \Theta\left[(\alpha + \Gamma G\mu) \tini - \ell - \Gamma G\mu
      \tsigma \right].
\end{aligned}
\label{eq:t4Fcur}
\end{equation}
The first term is the scaling loop distribution associated with the
Dirac LPF of equation~\eqref{eq:LPF-before}. The second term is the
initial distribution of loops at $\tini$ associated with the random
walk model of Vachaspati-Vilenkin~\cite{Vachaspati:1984}. Assuming the
random walk to be correlated over a length scale $\lcorr$, one
has~\cite{Vachaspati:1984}
\begin{equation}
  \Cini \simeq 0.4 \left(\frac{\tini}{\lcorr}\right)^{3/2}.
\label{eq:CiniVV}
\end{equation}
A natural value for $\lcorr$ is obtained by assuming that it is given by the
thermal process forming the strings, namely $\lcorr = 1/\Tini$. We
will, however, discuss various other possible choices in
section~\ref{sec:DM}.

At the time of condensation $\tcur$, the loops acquire quantum numbers
$N$, and we assume again a Dirac distribution for the generated charge:
\begin{equation}
  \Nn(\ell, \tcur, N) = \N(\ell, \tcur) \, \delta \negthinspace
  \left(N - \sqrt{\frac{\ell}{\lambda}}\right).
  \label{eq:cond}
\end{equation}
This is in agreement with Refs.~\cite{Brandenberger:1996zp,
  Peter:2013jj} and motivated by the fact that, if a thermal process
of temperature $\Tcur = 1/\lambda$ is at work during current
condensation, the conserved number $N$ laid down along the string
should be given by a stochastic process of root mean squared value
close to $\sqrt{\ell/\lambda}$.

String formation at $\tini$ and current condensation at $\tcur$ are
assumed to occur in the radiation era. In the following we will use as
model parameters $G\mu$ and $\calR$. The current condensation redshift
can be determined using entropy conservation:
\begin{equation}
  1 + \zcur = \left(\dfrac{\qcur}{\qzero} \right)^{1/3} \dfrac{\Tcur}{\Tcmb}\,,
\label{eq:zcurdef}
\end{equation}
where $\qcur=\q(\zcur)$, and $\qzero=\q(z=0)$, denotes the number of
entropic relativistic degrees of freedom at the time of current
condensation, and today, respectively. In the following, we consider
$\Tcur$ to be given by
\begin{equation}
  \Tcur = \dfrac{1}{\lambda} = \dfrac{\sqrt{\U}}{\calR}\,,
\end{equation}
and we take $\Tcmb = 2.725\,\uK$. In order to solve
equation~\eqref{eq:zcurdef} for $\zcur$, we have used the tabulated
values of $\q(z)$ associated with the thermal history in the Standard
Model and computed in Ref.~\cite{Hindmarsh:2005ix}. Still from entropy
conservation, the redshift associated with the formation of the string
network (at the temperature $\Tini$) is given by
\begin{equation}
1 + \zini =  \left(\dfrac{\qini}{\qzero} \right)^{1/3} \dfrac{\Tini}{\Tcmb}\,,
\end{equation}
where
\begin{equation}
\Tini = \sqrt{\U} = \calR \Tcur \,.
\end{equation}

\section{Cosmological distribution of vortons}
\label{sec:vorts}

From equation~\eqref{eq:sol1}, we can determine the distribution $\ud
\calN/\ud \cal \ell$ of relaxed vortons and produced vortons. Both of
these being stable, they will contribute to the relic content of the
universe.

Regarding the distributions of doomed loops and proto-vortons, these
could be important for some observational effects of strings, for
instance the stochastic gravitational wave background, but they cannot
contribute significantly to the dark matter content of the
Universe~\cite{Lorenz:2010sm}. Their distributions are determined from
equation~\eqref{eq:sol1} through
\begin{eqnarray}
\left. \N\right|_{\doom}(\ell, t) &\equiv& \int \ud N \Nn(\ell, t, N) \Theta(\calR - N),
       \label{eq:doomdef}
       \\
        \left.  \N\right|_{\proto}(\ell,t) &\equiv& \int \ud N \Theta(N - \calR) \Nn(\ell, t, N) \Theta\left[\ell - \ell_0(N)\right],
      \label{eq:protodef}
\end{eqnarray}
and are given in Appendix \ref{sec:others}.

In order to determine the vorton distribution, we recall that a vorton is a loop with topological number $N > \calR$ and size $\ell \leq \ell_0(N)$ if $\sigma > 0$.
In the limit $\sigma \rightarrow 0$, the charge $N$ of the vorton is proportional to its length $\ell_0(N) = N / \sqrt{\mu}$.  
In order to deal correctly with the singular behaviour in the limit $\sigma \rightarrow 0$, we firstly express the vorton distribution in terms of the charge $N$, namely calculate $\ud \calN / \ud N$, then take the limit $\sigma \rightarrow 0$, and finally determine $\ud \calN / \ud \ell$ through a simple change of variables since $\ell=\ell_0=N / \sqrt{\mu}$. 

Our starting point is therefore
\begin{equation}
    \left. \NN\right|_\vort(t, N) \equiv \Theta(N - \calR) \int \ud
    \ell \Nn(\ell, t, N) \Theta\left[\ell_0(N) - \ell\right],
    \label{eq:vortdef}
\end{equation}
which we calculate for both relaxed and produced vortons below.

\subsection{Relaxation term}

The distribution of the vortons coming from the initial conditions at the condensation is determined from \eqref{eq:vortdef}, substituting the first term of equation~\eqref{eq:sol1}, together with the initial distribution of loops in equation~\eqref{eq:cond}. This gives
\begin{equation}
    \left. \NN\right|_\rel = \Theta(N-\calR) \int_{-\infty}^{\ell_0(N)}  \left[\frac{a(\tcur)}{a(t)}\right]^3 \frac{\calJ(\lcur,N)}{\calJ(\ell, N)} \N\left(\lcur,\tcur\right) \delta\negthinspace\left(N - \sqrt{\frac{\lcur}{\lambda}}\right)\,\ud \ell,
\end{equation}
in which $\lcur(\ell,t,N)$, given in equation~(\ref{eq:lcur}), is the size of the loop at condensation.
In order to integrate over the Dirac delta distribution, we change integration variable from $\ell$ to
\begin{equation}
    y = N - \sqrt{\frac{\lcur}{\lambda}}\,,
\end{equation}
with corresponding Jacobian
\begin{equation}
     \frac{\ud y}{\ud \ell} = - \frac{1}{2 \sqrt{\lambda \lcur}} \left. \frac{\partial \lcur}{\partial \ell}\right|_{t,N} = - \frac{1}{2 \sqrt{\lambda \lcur}} \frac{\calJ(\lcur, N)}{\calJ(\ell, N)}\,,
\end{equation}
where we have used equation \eqref{eq:lcur}. As a result, the $\calJ$ terms cancel, and we obtain
\begin{equation}
    \left. \NN\right|_\rel = 2 \lambda N \Theta(N-\calR) \left[ \dfrac{a\left(\tcur \right)}{a(t)} \right]^3 \N\left(\lambda N^2,\tcur\right) \Theta\left\{\lcur[\ell_0(N), t, N] - \lambda N^2\right\}.
\end{equation}
In the limit $\sigma \rightarrow 0$,
the size of a vorton is $\ell = \ell_0(N)=N/\sqrt{\mu}$, and equation~\eqref{eq:lcur} simplifies to 
\begin{equation}
    \lcur[\ell_0(N), t, N] = \Gamma G\mu (t - \tcur) + \ell_0(N).
\end{equation}
Finally, using $\ud \calN/\ud \ell  = \sqrt{\mu} ~ \ud \calN/\ud N $, the vorton distribution generated from the initial loop distribution at $\tcur$ is given by
\begin{equation}
    \left. \N\right|_\rel(\ell, t) = 2 \lambda \mu \ell \left[ \frac{ a^3\left(\tcur\right)}{a^3(t) }\right] \N\left(\lambda \mu \ell^2,\tcur\right) \Theta\left[\Gamma G\mu (t - \tcur) + \ell - \lambda \mu \ell^2 \right] \Theta(\ell - \lambda).
    \label{eq:relaxed-vortons}
\end{equation}
This distribution scales like matter (modulo the time-dependence in
the $\Theta$-functions).  This term was already derived in
Ref.~\cite{Peter:2013jj}, and our results agree though the approach is
different.

We now turn to the vorton population sourced by loops chopped off from
the network, namely from the second term in equation~\eqref{eq:sol1}.

\subsection{Production term}

After the condensation, all the strings and loops carry a current,
which implies that all new loops formed from the network will inherit
the charge density carried by their mother strings. As a result, the
charged loop production function is still given by
equation~\eqref{eq:LPF-before}, modulated by the charge density
distribution, i.e.
\begin{equation}
    \calP(\ell,t,N) = C t^{-5} \delta\negthinspace\left(\frac{\ell}{t} - \alpha\right) \delta\negthinspace\left(N - \sqrt{\frac{\ell}{\lambda}}\right) \Theta(t - \tcur).
    \label{eq:lpf}
\end{equation}
Substituting into the last term of equation~\eqref{eq:sol1} (see~\cite{Auclair:2019jip} for more details) gives the number density
\begin{equation}
    \Nn(\ell,t,N) = \frac{C}{\calJ(\ell,N)}\left[\frac{a(\tstar)}{a(t)}\right]^3 \tstar^{-4} \frac{\calJ(\alpha \tstar,N)}{\alpha + \Gamma G\mu \calJ(\alpha \tstar,N)} \delta\negthinspace\left(N - \sqrt{\frac{\alpha \tstar}{\lambda}}\right) \Theta(\tstar - \tcur).
    \label{eq:d2NdlddN}
\end{equation}
where $\tstar(\ell,t,N)$ is the time of loop formation, obtained by
solving
\begin{equation}
    \Gamma G\mu \tstar + \xi(\alpha \tstar,N) = \Gamma G\mu t + \xi(\ell,N),
    \label{eq:tstarGen}
\end{equation}
which again follows from the fact that $2v = \Gamma G\mu t +
\xi(\ell,N)$ is a conserved quantity during the lifetime of the loops.
The definition in equation~\eqref{eq:vortdef} then gives
\begin{equation}
\begin{aligned}
  \left. \frac{\ud \calN}{\ud N}\right|_\pro \negthickspace \negthickspace\negthickspace &= \Theta\left(N -
  \lambda\sqrt{\mu}\right) \\ & \times  \int_{-\infty}^{\ell_0(N)} \ud \ell \frac{C}{\calJ(\ell,N)}\left[\frac{a(\tstar)}{a(t)}\right]^3 \tstar^{-4} \frac{\calJ(\alpha \tstar,N)}{\alpha + \Gamma G\mu \calJ(\alpha \tstar,N)} \delta\negthinspace\left(N - \sqrt{\frac{\alpha \tstar}{\lambda}}\right) \Theta(\tstar - \tcur).
    \label{vor}
\end{aligned}
\end{equation}
We again integrate the Dirac delta distribution by means of the change of variable
\begin{equation}
    \tilde{y} = N - \sqrt{\frac{\alpha \tstar}{\lambda}}\,,
\end{equation}
with corresponding Jacobian
\begin{equation}
    \frac{\ud \tilde{y}}{\ud \ell} = - \sqrt{\frac{\alpha}{\lambda}} \frac{1}{2\sqrt{\tstar}}\left. \frac{\partial \tstar}{\partial \ell}\right|_{t,N} = - \sqrt{\frac{\alpha}{\lambda}} \frac{1}{2\sqrt{\tstar}} \frac{\calJ(\alpha \tstar, N)}{\calJ(\ell, N)[\alpha + \Gamma G\mu \calJ(\alpha \tstar)]} \,.
\end{equation}
Thus equation~\eqref{vor} gives
\begin{equation}
\begin{aligned}
  \left. \frac{\ud \calN}{\ud N}\right|_\pro & = \Theta\left(N -
  \lambda\sqrt{\mu}\right) \\ & \times \frac{2 \lambda N}{\alpha} C \left[\frac{a\left(\lambda N^2/\alpha\right)}{a(t)}\right]^3 \left(\frac{\lambda N^2}{\alpha}\right)^{-4}
    \Theta(\lambda N^2 - \alpha \tcur) \Theta\left[\tstar(\ell_0(N), t, N) - \frac{\lambda N^2}{\alpha}  \right].
\end{aligned}
\end{equation}
In the limit $\sigma \rightarrow 0$, equation~(\ref{eq:tstarGen}) reduces to
\begin{equation}
    (\alpha + \Gamma G\mu) \tstar = \ell_0(N) + \Gamma G\mu t,
    \label{eq:tstarNG}
\end{equation}
and, using the fact that vortons have size $\ell = \ell_0(N) =
N/\sqrt{\mu}$, it follows that the produced vorton distribution is
given by
\begin{equation}
\begin{aligned}
  \left.\N\right|_\pro & =  \frac{2 \lambda \mu \ell}{\alpha} C
  \left[\frac{a\left(\lambda \mu \ell^2/\alpha\right)}{a(t)}\right]^3
  \\ & \times \left(\dfrac{\lambda \mu \ell^2}{\alpha}\right)^{-4}
    \Theta(\lambda \mu \ell^2 - \alpha \tcur) \Theta\left(\frac{\Gamma G\mu t + \ell}{\alpha + \Gamma G\mu} - \frac{\lambda \mu \ell^2}{\alpha} \right)\Theta\left(\ell - \lambda\right),
    \label{eq:produced-vortons}
\end{aligned}
\end{equation}
which again scales as matter.

\begin{figure}
    \centering
    \begin{subfigure}{0.49\textwidth}
    \includegraphics[width=\textwidth]{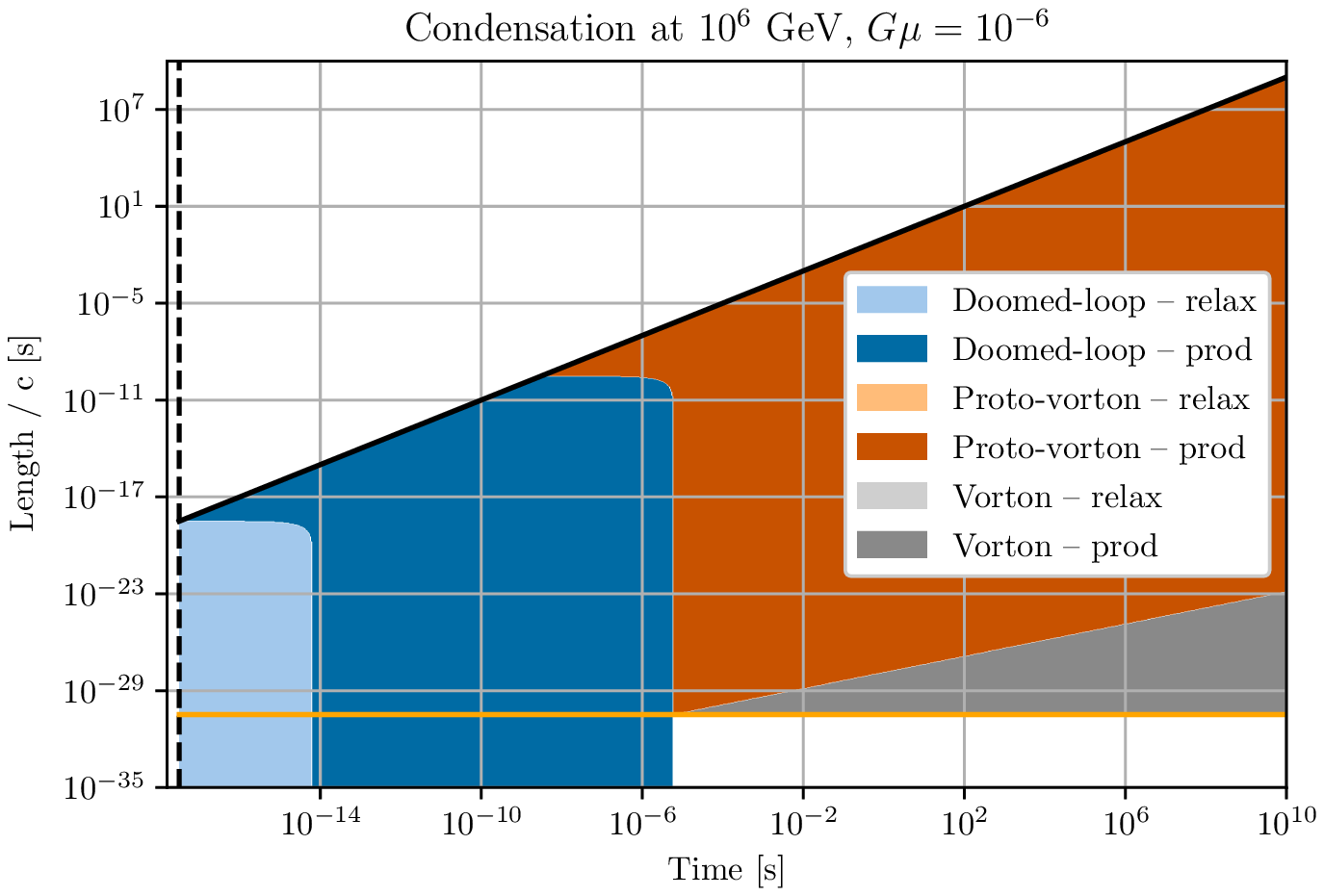}
    \end{subfigure}
    \begin{subfigure}{0.49\textwidth}
    \includegraphics[width=\textwidth]{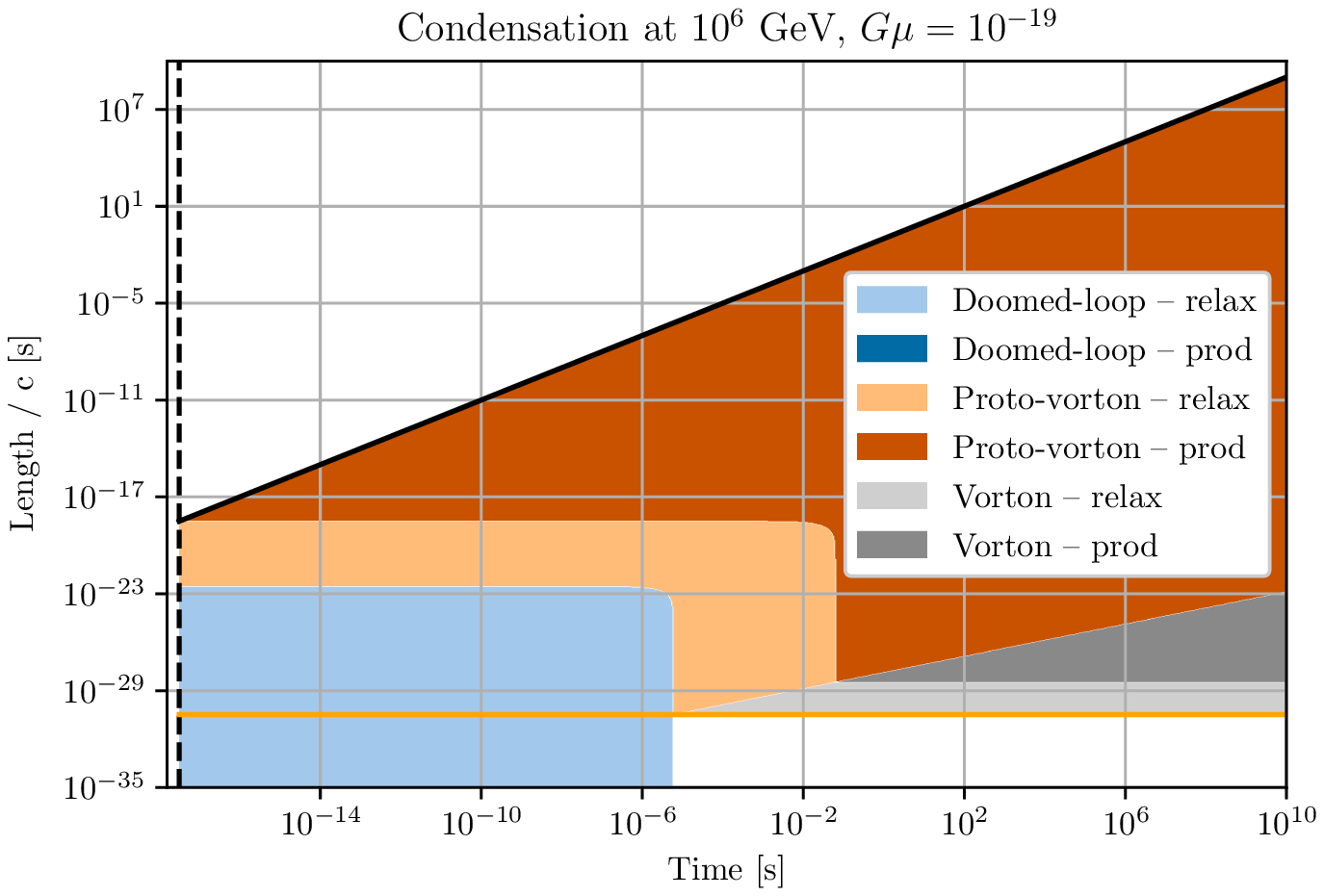}
    \end{subfigure}
    \caption{Diagram $(\ell, t)$ for the different types
      loops/vortons. The left panel is for $G\U = 10^{-16}$ and the
      right panel for $G\U=10^{-19}$. The dark-dashed vertical line is
      the time of condensation, when strings become superconducting.
      The diagonal dark line represents $\ell = \alpha t$ (with $\alpha = 0.1$) the size at
      which loops are produced.
      The orange horizontal line shows the value of $\lambda$.}
    \label{fig:regions}
\end{figure}

In figure~\ref{fig:regions} we show the different regions of
$(\ell,t)$-space which are populated by either relaxed or
produced vortons, and also proto-vortons and doomed loops (see
Appendix \ref{sec:others}).  Essentially, for vortons, these are fixed
by the $\Theta$-functions in equation~(\ref{eq:produced-vortons}) and
equation~(\ref{eq:relaxed-vortons}). In particular we observe that for
\begin{equation}
    G\mu > \dfrac{\alpha G \tcur}{\lambda^3} \iff \calR > \dfrac{\alpha \tcur}{\lambda}\,,
\end{equation}
there are no relaxed vortons produced, explaining the differences
between the two panels of figure~\ref{fig:regions}.

A consequence of the different $\Theta$-functions in
equation~(\ref{eq:produced-vortons}) is that when evaluating $\tstar$, the
formation time of loops, it turns out that all vortons were produced
initially during radiation era.  If one imposes that the loop
production function of equation~\eqref{eq:lpf} is only valid for $t <
\teq$, one finds that equation~\eqref{eq:produced-vortons} is
multiplied by the Heaviside function $\Theta(\alpha \teq - \lambda \mu
\ell^2)$.

\section{Relic abundance}
\label{sec:DM}

In the previous sections we have established that the number density
of vortons produced during the radiation era contains two components,
namely the relaxed vortons with length distribution given in
equation~\eqref{eq:relaxed-vortons}, and the produced vortons with
length distribution given in equation~\eqref{eq:produced-vortons}.

\subsection{Analytic estimates}
\label{sec:analytics}

In order to estimate the density parameter associated with the relic
vortons today, we can use the results of the previous section
evaluated at present time $t=\tzero$. The density parameter for each
population is defined by
\begin{equation}
  \Omega \equiv \dfrac{8\pi G\U}{3 \Hzero^2} \int_0^\infty \ell
  \N(\ell,\tzero)\, \ud \ell.
\label{eq:Omegadef}
\end{equation}
Starting with the contribution of the relaxed vortons, from
equation~\eqref{eq:relaxed-vortons}, estimated today, the
dimensionless loop distribution reads
\begin{equation}
\tzero^4 \left.\N\right|_{\rel} = \dfrac{2\calR^2}{(1+\zcur)^3}
\dfrac{\ell}{\lambda} \left(\dfrac{\tzero}{\tcur} \right)^4 \tcur^4 \N(\calR^2 \ell^2/\lambda,\tcur)
\heaviside{\ell-\lambda} \heavisideb{\ellT(\tzero)-\ell},
\label{eq:t4FiniToday}
\end{equation}
where we have introduced the typical length~\cite{Peter:2013jj}
\begin{equation}
\ellT(\tzero) \equiv \dfrac{\lambda}{2 \NNstar^2} \left[1 +
  \sqrt{1+4\NNstar^2 \dfrac{\gammad(\tzero-\tcur)}{\lambda}} \right],
\label{eq:ellT}
\end{equation}
solution of the quadratic equation appearing in the argument of the
first Heaviside function in equation~\eqref{eq:relaxed-vortons}. As
explicit in the above expression, this is the maximal possible length
of a relaxed vorton today, larger loops belonging to the (relaxed)
proto-vorton distribution, see also figure \ref{fig:regions}. In this
expression, the loop distribution at $\tcur$ is given by
equation~\eqref{eq:t4Fcur}. The vorton distribution of
equation~\eqref{eq:t4FiniToday} obtained by taking, in
equation~\eqref{eq:t4Fcur}, $C=0$ and $\Cini$ given by
equation~\eqref{eq:CiniVV} is the one originally considered and
derived in Ref.~\cite{Brandenberger:1996zp}. We see that by
considering $C\ne0$, i.e. by including all the Nambu-Goto loops
produced between $\tini$ and $\tcur$, we are adding a new population,
not considered so far, to the relaxed vorton abundance.

It is actually possible to derive an analytical expression for the
density parameter of these new relaxed vortons only. Let us consider a
loop distribution at $\tcur$ given by equation~\eqref{eq:t4Fcur} with
$C\ne0$ and $\Cini=0$. In other words, we take the extreme situation
in which at $t=\tini$, there is no loop at all. All loops present at
$\tcur$ are therefore created from the network between $\tini$ and
$\tcur$. Plugging equation \eqref{eq:t4FiniToday} into
\eqref{eq:Omegadef}, one gets after some algebra
\begin{equation}
\begin{aligned}
  \Omegarelmin &=\dfrac{2 \NNstar^2 C}{9\left(1+\zcur\right)^3  (\Hzero \tcur)^2}
  \dfrac{\left(\alpha + \gammad
    \right)^{3/2}}{ (\Mp \tcur)^2 \gammad } \\ & \times \left\{
  \!\dfrac{\xmax^3}{\left[\gammad + \left(\lambda
      \xmax/\ellbarcur\right)^2 \right]^{3/2}} - \dfrac{\xmin^3}{\left[\gammad + \left(\lambda
      \xmin/\ellbarcur\right)^2 \right]^{3/2}} \!\right\},
\end{aligned}
\label{eq:omegarelmin}
\end{equation}
with the dimensionless numbers
\begin{equation}
  \xmax \equiv
  \min\negthinspace\left(\dfrac{\ellT}{\lambda},\dfrac{\ellbarcur}{\lambda}
  \right), \qquad \xmin = \max\negthinspace\left[1,\dfrac{1}{\NNstar}
  \sqrt{\dfrac{\ellbarini(\tcur)}{\lambda}} \right],
\end{equation}
and where we have introduced the new length scales
\begin{equation}
\ellbarcur \equiv \dfrac{\sqrt{\alpha \lambda \tcur}}{\NNstar}\,,
\qquad \ellbarini(\tcur) \equiv \tini \left(\alpha + \gammad \right) -
\gammad \tcur\,.
\end{equation}
From the fact that we started with no loop at all at the string
forming time $\tini$, equation~\eqref{eq:omegarelmin} is necessarily a
robust lower bound for the relaxed vorton abundance today. These
objects will be referred to as the ``irreducible relaxed vortons''.

\begin{figure}
    \centering
    \includegraphics[width=\twofigw]{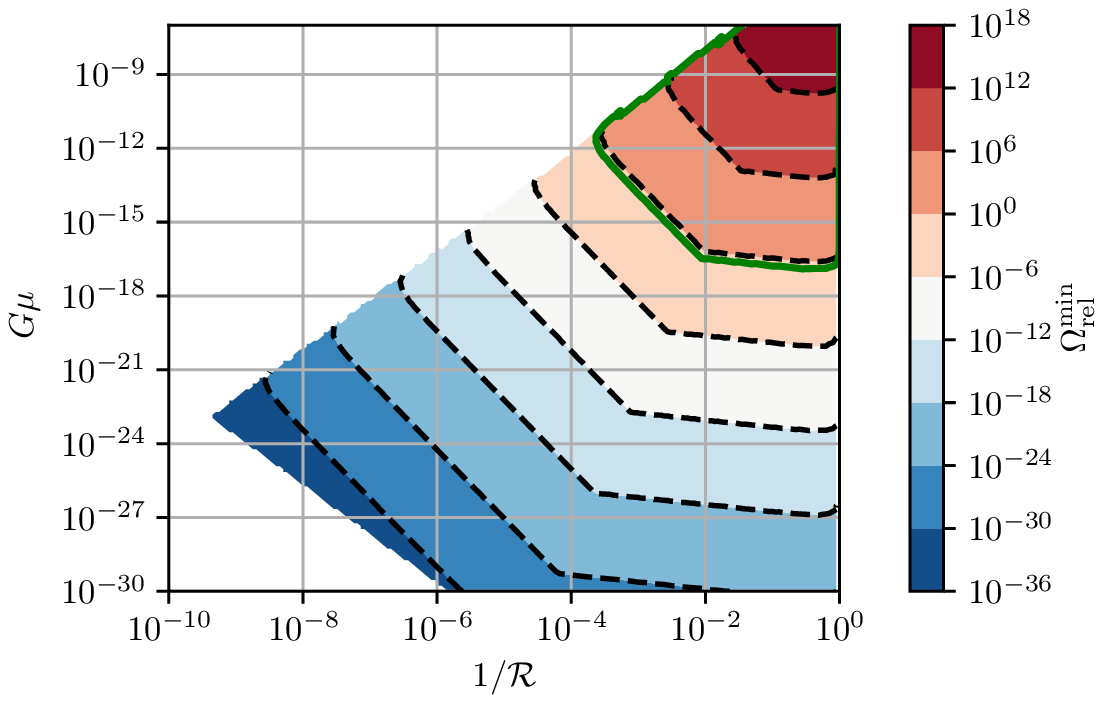}
    \includegraphics[width=\twofigw]{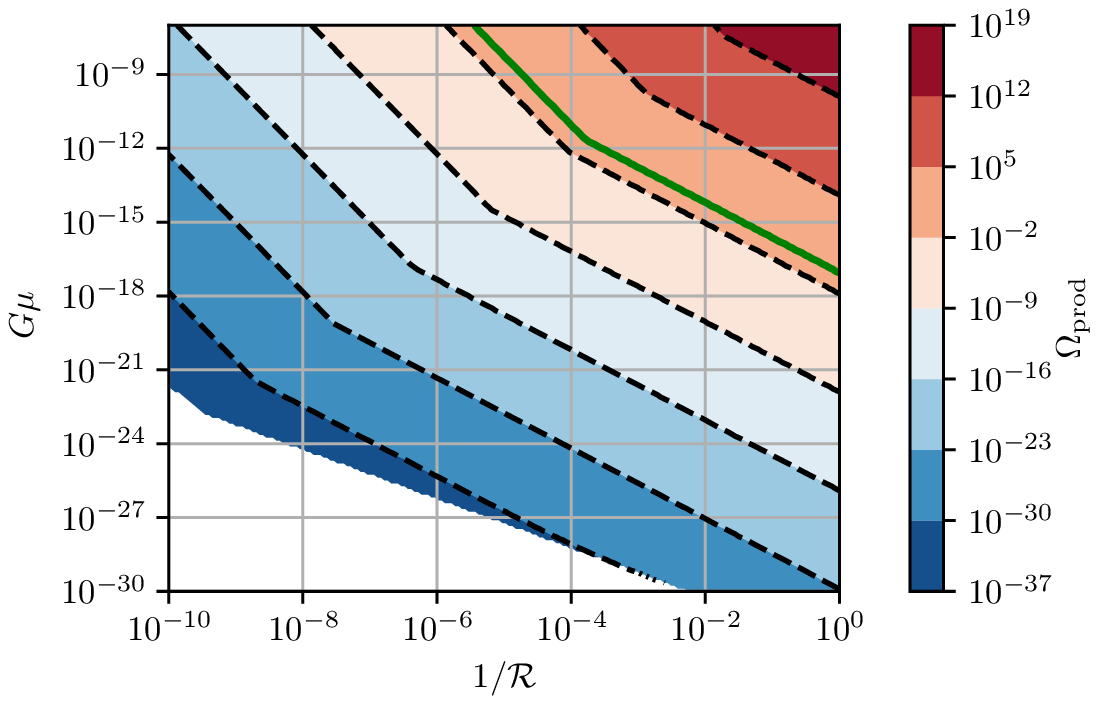}
    \caption{The left panel shows the density parameter $\Omegarelmin$
      (today) from the population of \emph{irreducible relaxed}
      vortons, i.e.~we have assumed that there is no loop at the
      string forming time ($\Cini=0$). The right panel shows the
      density parameter $\Omegaprod$ of \emph{produced} vortons
      derived analytically in equation~\eqref{eq:omegaprodapprox}. The
      thick green line shows the value $\OmegaDM=0.3$, typical of the
      current dark matter density parameter. The white patches on
      these figures correspond to regions of the parameter space where
      no vortons are present: all loops there are either doomed or
      proto-vortons. Abundances of these two populations of vortons
      have not been derived before and constitute an irreducible contribution.}
    \label{fig:omegaapprox}
\end{figure}

Similarly, the produced vorton density distribution today is given by
equation~\eqref{eq:produced-vortons} evaluated at $t=\tzero$. The
dimensionless distribution today reads
\begin{equation}
\tzero^4 \left.\N\right|_{\pro} = \dfrac{2 C}{\left[1 +
    z(\tell)\right]^3} \left(\dfrac{\alpha}{\NNstar}
\dfrac{\lambda}{\tzero} \right)^3 \left(\dfrac{\tzero}{\ell} \right)^7
\heaviside{\ell - \ellbarcur} \heavisideb{\ellbarT(\tzero) - \ell} \heaviside{\ell-\lambda},
\label{eq:t4FprodToday}
\end{equation}
where we have made explicit the new length scale
\begin{equation}
\ellbarT(\tzero) \equiv \dfrac{\lambda}{2\NNstar^2}
\dfrac{\alpha}{\alpha + \gammad} \left[1 + \sqrt{1+ 4 \NNstar^2
    \dfrac{\alpha + \gammad}{\alpha} \dfrac{\gammad \tzero}{\lambda}}
  \right],
\end{equation}
which is the analogue of $\ellT(\tzero)$ but for the produced vortons,
see equation~\eqref{eq:ellT}. This is the maximal possible size of a
produced vorton today. Let us notice the appearance of the redshift
$z(\tell)$, evaluated at some (past) $\ell$-dependent cosmic time
\begin{equation}
\tell \equiv \dfrac{\NNstar^2}{\alpha} \dfrac {\ell^2}{\lambda}\,.
\label{eq:tell}
\end{equation}
Plugging equation~\eqref{eq:t4FprodToday} into \eqref{eq:Omegadef}, one gets
\begin{equation}
\begin{aligned}
  \Omegaprod = \dfrac{16 \pi G \U}{3 (\Hzero \tzero)^2} C
  \left(\dfrac{\alpha}{\NNstar^2} \right)^3
  \left(\dfrac{\tzero}{\lambda} \right)^2 \int_{\ymin}^{\ymax}
  \dfrac{\left[1 + z(\ty)\right]^3}{y^6} \ud y\,,
\end{aligned}
\label{eq:omegaprodexact}
\end{equation}
with
\begin{equation}
\ymin \equiv \max\negthinspace\left(1,\dfrac{\ellbarcur}{\lambda}\right),
\qquad \ymax \equiv \dfrac{\ellbarT(\tzero)}{\lambda}\,.
\end{equation}
Equation~\eqref{eq:omegaprodexact} shows that the knowledge of the
whole thermal history of the Universe through $z(\ty)$ is a priori
required to accurately determine $\Omegaprod$. This is expected as the
``time of flight'' of a proto-vorton between its creation and
stabilisation as a vorton depends on its size at formation. Therefore,
at any given time, the population of produced vortons keeps a
memory of the past history of the Universe.

The integral~\eqref{eq:omegaprodexact} can be analytically performed
with some simplifying assumptions. One can consider an exact power-law
expansion for the radiation and matter era together with an
instantaneous transition at $\teq$. Taking $a(t) \propto t^{\nu}$,
with $\nu=\nurad\equiv1/2$ and $\nu=\numat\equiv 2/3$ in the radiation
and matter era, respectively, one gets
\begin{equation}
\begin{aligned}
  \Omegaprod & =  \dfrac{16 \pi G \U}{3 (\Hzero \tzero)^2} C
  \left(\dfrac{\alpha}{\NNstar^2} \right)^3
  \left(\dfrac{\tzero}{\lambda} \right)^{2-3\numat}  \\ & \times \left\{  
  \left(\dfrac{\NNstar^2}{\alpha}\right)^{3\nurad}
  \dfrac{\left[\min\left(\ymax,\yeq \right)\right]^{6\nurad-5} -
    \ymin^{6\nurad-5}}{5 - 6\nurad}
  \left(\dfrac{\teq}{\lambda}\right)^{3(\numat-\nurad)} \right.
    \\ & + \left.  \left(\dfrac{\NNstar^2}{\alpha}\right)^{3\numat} \dfrac{\ymax^{6\numat-5}
    - \left[\max\left(1,\yeq\right)\right]^{6\numat-5}}{5-6\numat} \right\},
\end{aligned}
\label{eq:omegaprodapprox}
\end{equation}
where
\begin{equation}
\yeq \equiv \dfrac{\ellbareq}{\lambda}\,, \quad \textrm{with} \quad
\ellbareq \equiv \dfrac{\sqrt{\alpha \lambda \teq}}{\NNstar}\,.
\end{equation}
Unsurprisingly, the particular cosmic time $\teq$ imprints a new
length scale $\ellbareq$ in the distribution.

We have represented in figure~\ref{fig:omegaapprox} both
$\Omegarelmin$ and $\Omegaprod$ as a function of $(G\U,1/\NNstar)$
given by the equations~\eqref{eq:omegarelmin} and
\eqref{eq:omegaprodapprox}. The thick green line shows the contour
matching the value $\OmegaDM=0.3$. For the irreducible relaxed
vortons, the only additional parameter entering
equation~\eqref{eq:omegarelmin} is $\zcur$, which has been determined
using $a(t)\propto t^{\nurad}$ for the radiation era together with the
thermal initial conditions of equation~\eqref{eq:zcurdef} (using
$\qcur=104$). As already discussed, these two populations of vortons
are an unavoidable consequence of the loop production associated with
a scaling cosmic string network and have not been considered
before. For instance, taking $\NNstar \lesssim 10^{2}$, these figures
show that all values of $G\U$ greater than $10^{-15}$ are overclosing
the Universe with vortons, even though no loops at all are present at
$\tini$ when the strings are formed. Although not very visible on the
figure, there is a small region around $\NNstar=1$ in which
$\Omegarelmin=0$. Indeed, if $\tini=\tcur$ and $\Cini=0$, there is no
time at all to produce loops before the current appears. However, this
region is actually ruled out as filled with vortons produced
afterwards (see right panel of figure~\ref{fig:omegaapprox}).

Returning to equations~\eqref{eq:t4FiniToday} and \eqref{eq:t4Fcur}, the
most general situation for the relaxed vortons is to start with a
mixture of loops created at the string forming time and loops created
from the network between $\tini$ and $\tcur$, i.e.~one has both
$C\ne0$ and $\Cini \ne 0$. Moreover, from
equation~\eqref{eq:omegaprodexact}, the accurate expression for
$\Omegaprod$ requires specifying the whole thermal history of the
Universe and the integral has to be performed numerically. In the next
section, we numerically integrate both $\Omegarel$ and $\Omegaprod$
and discuss their sensitivity to the initial conditions.

\subsection{Numerical integration and initial conditions}
\label{sec:numerics}

\begin{figure}
\begin{center}
  \includegraphics[width=\twofigw]{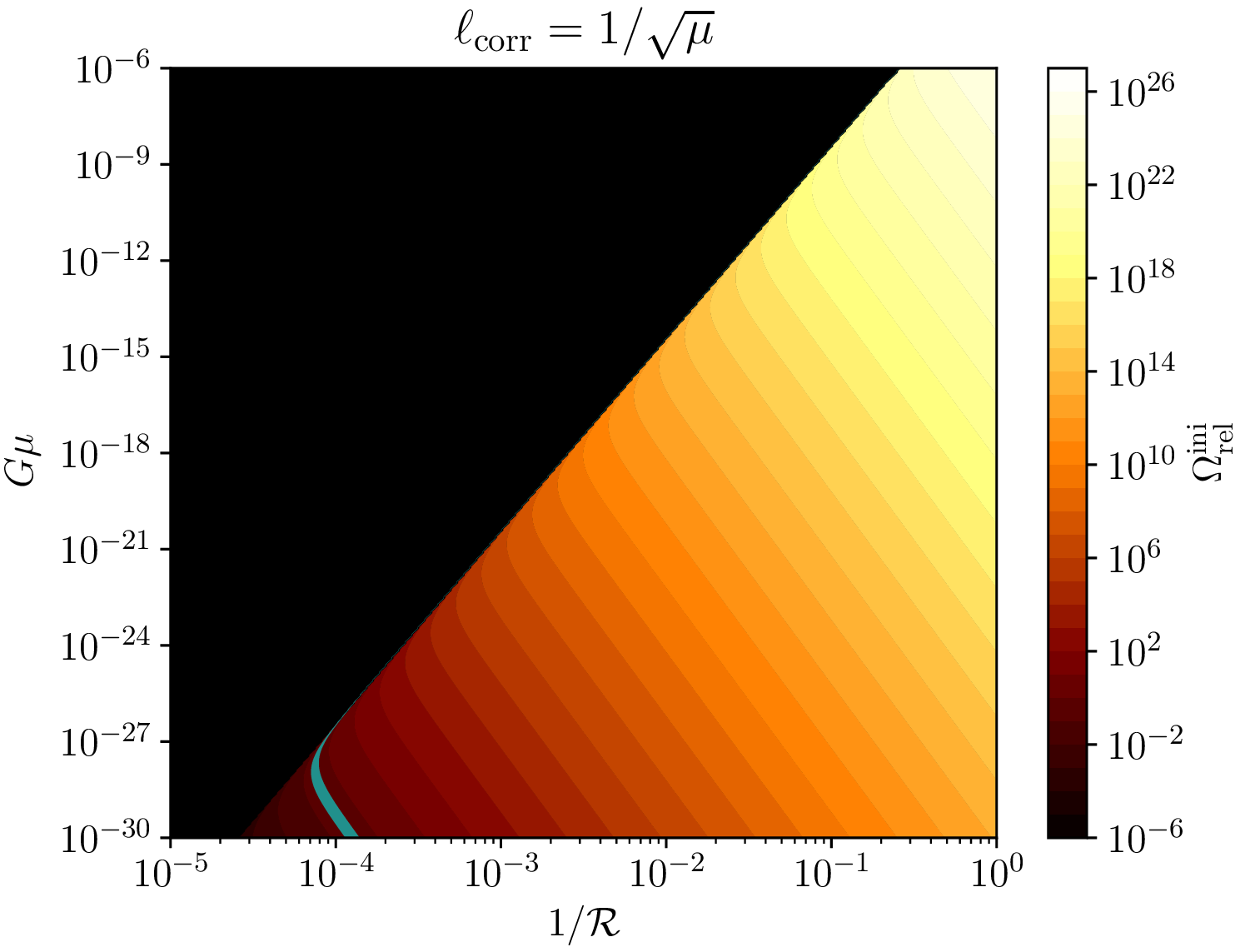}
  \includegraphics[width=\twofigw]{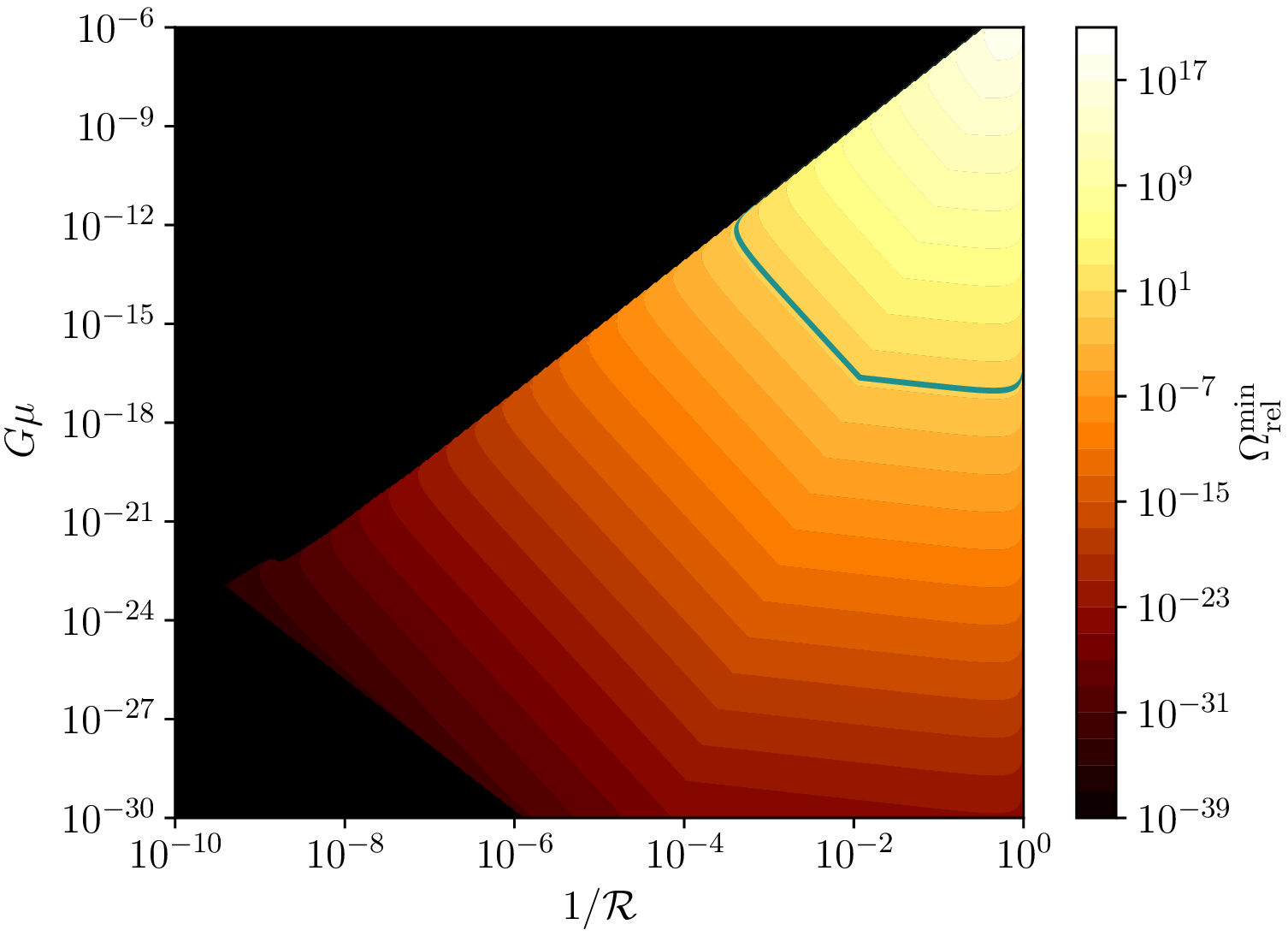}
  \includegraphics[width=\twofigw]{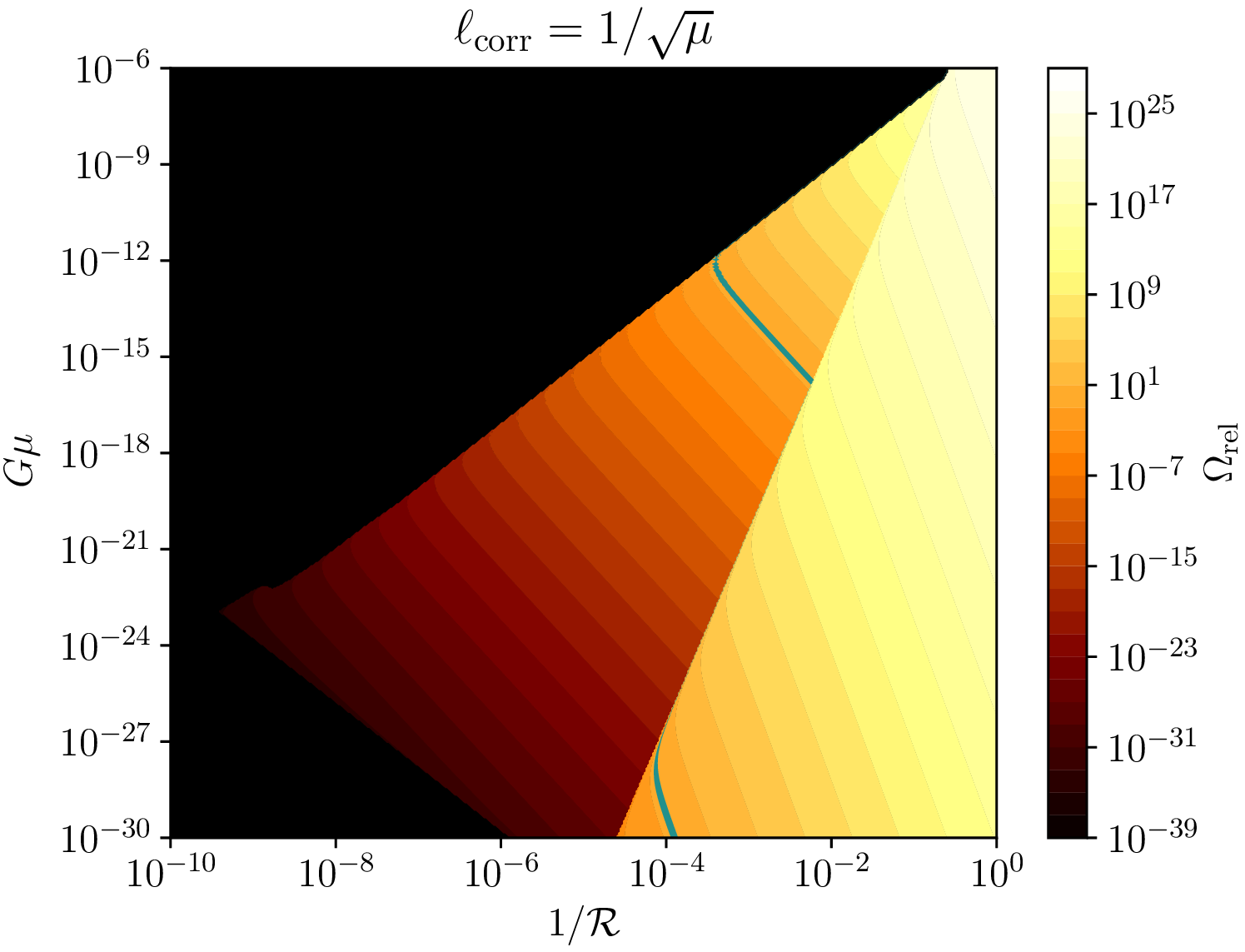}
  \includegraphics[width=\twofigw]{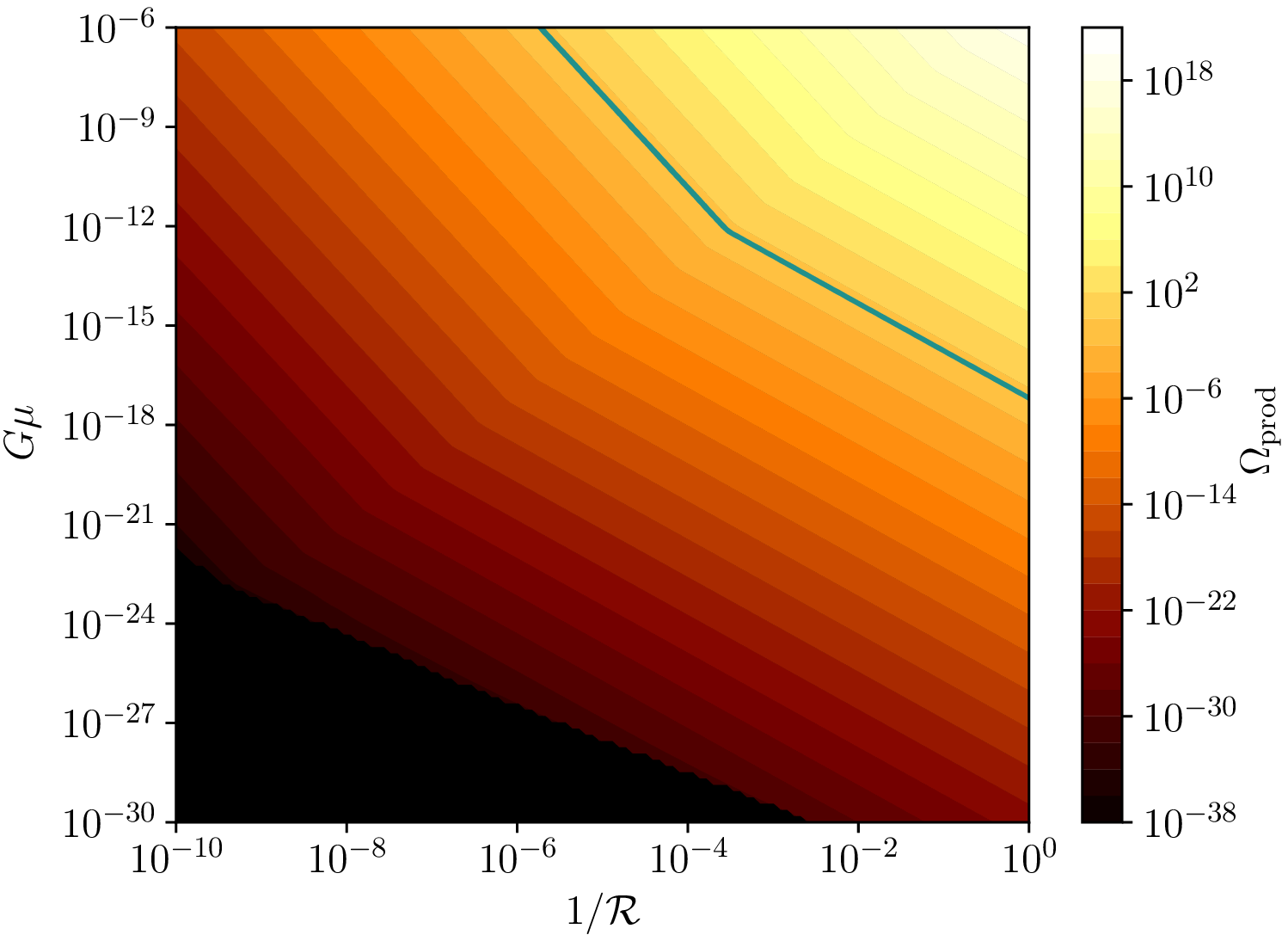}
  \caption{The upper left-hand panel shows the density parameter of relaxed
    vortons coming only from loops present at the string-forming phase
    transition, when starting from a Vachaspati-Vilenkin distribution
    at $t=\tini$. This is the population derived in
    Ref.~\cite{Brandenberger:1996zp}, that we recover by setting
    $C=0$ in our equations. The upper right-hand panel shows the
    numerically evaluated density parameter of the irreducible relaxed
    vortons $\Omegarel^{\min}$ (to be compared to our analytic
    estimation in the left panel of figure~\ref{fig:omegaapprox}). The
    lower left-hand panel shows the density parameter $\Omegarel$ (today)
    from the population of all \emph{relaxed} vortons (the sum of the
    upper left and right panels). Thermal history effects are visible
    on the upper boundary towards the minimum possible values of
    $1/\NNstar$ and $G\U$. The lower right-hand panel shows the density
    parameter $\Omegaprod$ today of \emph{produced} vortons derived
    numerically, and is indistinguishable from our analytic estimation
    of equation~\eqref{eq:omegaprodapprox} (see right-hand panel of
    figure~\ref{fig:omegaapprox}). The thick green line corresponds to
    all density parameter values in the range $[0.2,0.4]$.}
\label{fig:omegasvv}
\end{center}
\end{figure}

\begin{figure}
\begin{center}
  \includegraphics[width=\onefigw]{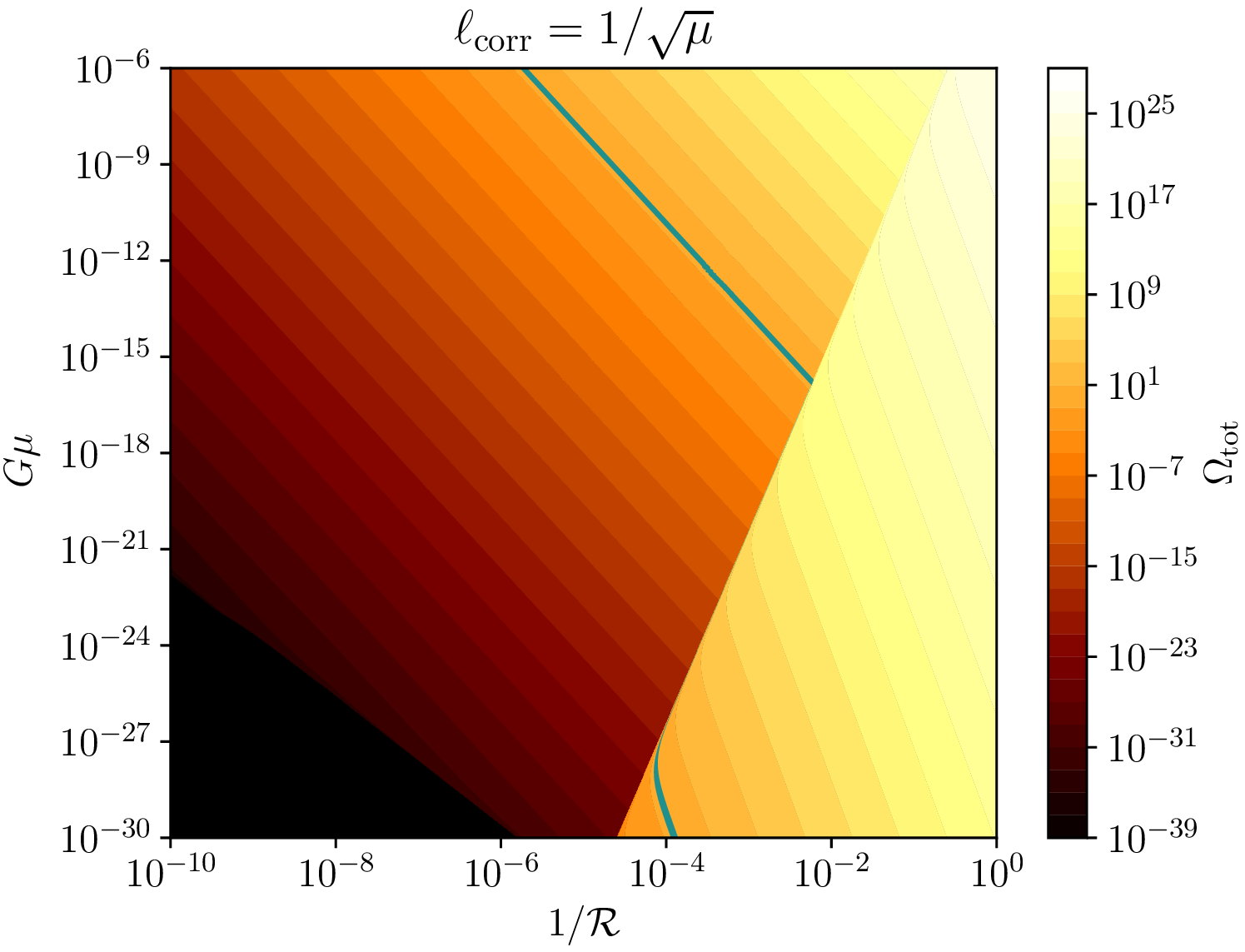}
\caption{The total relic abundance of all vortons starting from a
  Vachaspati-Vilenkin initial loop distribution, with an initial
  thermal correlation length $\ellcorr = 1/\sqrt{\U}$, and a one-scale
  loop production function with $\alpha=0.1$. The green line
  corresponds to the range of values $[0.2,0.4]$. The different
  populations contribution is represented in
  figure~\ref{fig:omegasvv}.}
\label{fig:omegatotvv}
\end{center}
\end{figure}

Compared to the previous section, we now numerically integrate both
$\Omegarel$ and $\Omegaprod$ starting from the general initial loop
distribution described in section~\ref{sec:icloop}. Thermal initial
conditions are taken assuming that the number of relativistic degrees
of freedom is given by the Standard Model as derived in
Ref.~\cite{Hindmarsh:2005ix}.

Figures \ref{fig:omegasvv} and \ref{fig:omegatotvv} show the density
parameters today of all the relaxed vortons, the produced vortons and
the sum of the two contributions when the string forming network at
$t=\tini$ is given by the Vachaspati-Vilenkin initial condition (see
section~\ref{sec:icloop}). This implies that the typical size of loops
at $\tini$ is given by thermal fluctuations of the Higgs field and
$\ellcorr = 1/\sqrt{\U}$.

The lower right panel of figure~\ref{fig:omegasvv}, compared to the
right panel of figure~\ref{fig:omegaapprox}, shows that our
approximated formula~\eqref{eq:omegaprodapprox} is relatively
accurate. The lower left panel of figure~\ref{fig:omegasvv} exhibits a
triangle-like region which is not visible on the left panel of
figure~\ref{fig:omegaapprox}. This region, with a high density of
relaxed vortons, is precisely the one associated with the relaxed
vortons created from the loops initially present at the string forming
time and which were studied in Ref.~\cite{Brandenberger:1996zp}. This
contribution is represented alone in the upper left panel of
figure~\ref{fig:omegasvv}. In this corner of parameter space, we
recover the results already presented in
Ref.~\cite{Brandenberger:1996zp}: essentially all values of $G \U$ are
ruled out, only values of $G\U = \order{10^{-30}}$ and
$\NNstar=\order{10^{4}}$ remain compatible with the cosmological
bounds.

When all contributions are combined, as shown in 
figure~\ref{fig:omegatotvv}, one can see
that for all $G\U$ there are
values of $\NNstar$ which make the vortons either an acceptable
candidate for dark matter (green line) or a subdominant component
today (left of the green line). However, this figure also
shows that there is an absolute lower bound for $\NNstar$ below which
vortons would overclose the universe, independently of the value of $G
\U$ (which is also given by the green line). For instance, there are no
acceptable regions for which $\NNstar < 10^2$, implying that stable
vortons in our Universe can only be created if the temperature of
current condensation is {\it at least} two orders of magnitude lower than
the one of the formation of strings. This result is the consequence of
the irreducible relaxed and produced vorton contributions closing the
parameter space up to the maximum admissible values of $G\U$. It may
have some implications on the particle physics models creating strings
and currents~\cite{2003PhRvD..68j3514J, 2005JCAP...03..004R}.

\begin{figure}
\begin{center}
  \includegraphics[width=\onefigw]{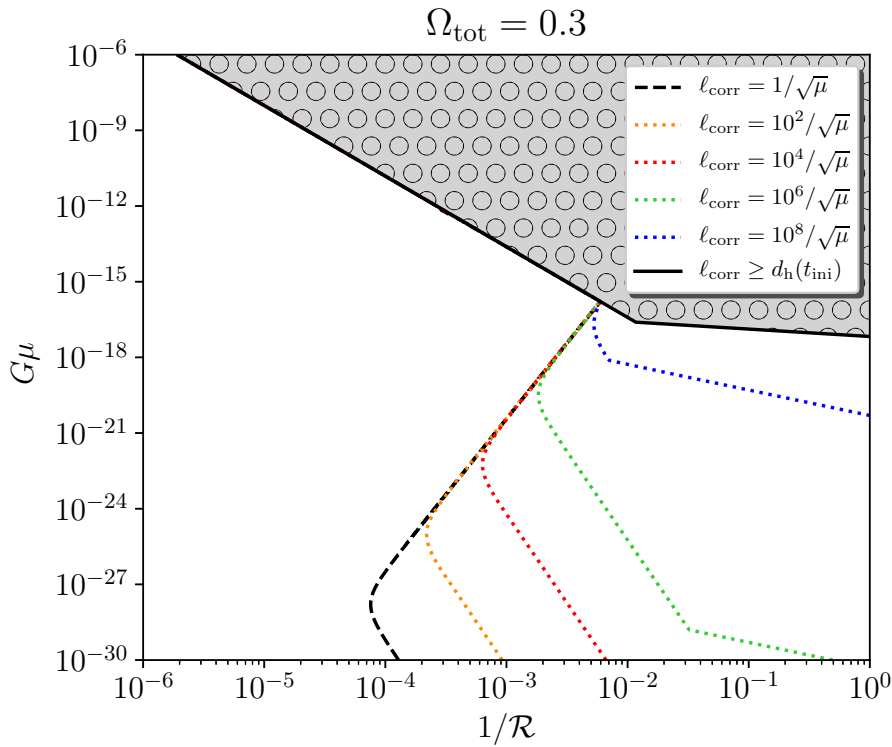}
\caption{The total relic abundance of all vortons starting from a
  Vachaspati-Vilenkin initial loop distribution with various
  correlation length $\ellcorr$ ranging from the thermal one
  $1/\sqrt{\U}$ to the Kibble one $\horizon{\tini}$. Each curve
  represents the value $\Omegatot=0.3$. Domains right of this curve
  lead to vortons overclosing the Universe, domains on the left are
  compatible with current cosmological constraints. The upper hatched
  region corresponds to the irreducible relaxed and produced vortons
  not affected by the initial conditions.}
\label{fig:depcorr}
\end{center}
\end{figure}
Despite the fact that Vachaspati-Vilenkin initial conditions are quite motivated
from the point of view of a thermal process, loops could be created
from other processes~\cite{Rajantie:2001ps, Rivers:2002vm}. Therefore,
instead of assuming $\ellcorr=1/\sqrt{\U}$, one could use the Kibble
argument~\cite{Kibble:1976sj, Kibble:1981gv} and take $\ellcorr =
\horizon{\tini}$, where $\horizon{\tini} = 2 \tini$ denotes the distance to the would-be
particle horizon at the string forming time. Doing so leads to the
same overall relic abundance of vortons as in
section~\ref{sec:analytics} where we were assuming $\Cini=0$. There
are simply not enough loops initially, compared to the one produced
later on, to significantly change the final density parameter.

In order to quantitatively study the dependence of $\Omegatot$ with
respect to the loop distribution at $\tini$, we have represented in
figure~\ref{fig:depcorr} the values of $\Omegatot=0.3$ in the plane
$(G\U,1/\NNstar)$ for various choices of $\ellcorr$. They range from
the thermal value $\ellcorr = 1/\sqrt{\U}$ to the causal one
$\ellcorr=\horizon{\tini}$, and even above, a situation that could
appear if loops have been formed during cosmic
inflation~\cite{2016JCAP...02..033R}. Everything on the right of the
lines represented in this figure would lead to an overclosure of the
Universe, while everything on the left is compatible with current
measurements. The hatched region in this figure shows the robust bound
discussed earlier, where there are only irreducible relaxed vortons
and produced vortons.

In all our analysis and equations, we have left the parameter $\alpha$ arbitrary, fixing only $\alpha=0.1$ for the figures for well motivated reasons. Changing $\alpha$ to smaller values, while keeping everything else fixed, increases the population of doomed loops, and thus decreases the vortons abundance. The explicit dependence in $\alpha$ can be read off from equations~\eqref{eq:omegarelmin} and \eqref{eq:omegaprodapprox}.

\subsection{Other observables}
\label{sec:observables}

A network of cosmic strings can let imprints in various cosmological
observables, such as the stochastic background of gravitational waves
and the Cosmic Microwave Background (CMB). In the present case, the
stabilisation of vortons is expected to prevent a part of the energy
to be converted into gravitational waves. We have therefore estimated
the gravitational wave power spectrum generated from proto-vortons and
doomed loops only. Their loop number densities are explicited in the
appendix~\ref{sec:others}. Due to the very small size of the vortons,
the lack of energy in terms of gravitational waves ends up being
negligible and the predictions for the stochastic background of
gravitational waves remain unchanged compared to Nambu-Goto strings
with a one-scale loop production function~\cite{Auclair:2020oww}. For
the one-scale LPF, the current Laser Interferometer Gravitational-Wave
Observatory (LIGO) bound on the string tension is $G \U <
\order{10^{-11}}$~\cite{Blanco-Pillado:2017rnf, Ringeval:2017eww,
  Abbott:2017mem} but depends on some assumptions on the string
microstructure. Concerning the CMB, detectable distortions induced by
cosmic strings are mostly due to the long strings in scaling such that
they are not sensitive to the loop distribution and provide a robust
upper bound $G\U < \order{10^{-7}}$ for all types of
strings~\cite{Ringeval:2012tk, Ade:2013xla, Lizarraga:2014xza,
  Lazanu:2014eya, Lazanu:2014xxa}. Both of these bounds therefore
apply to current-carrying strings with vortons. Let us also remark
that current-carrying strings may lead to other observational
signatures, for instance gamma ray or radio bursts~\cite{Cai:2011bi,
  Cai:2012zd, Auclair:2019jip}.

\section{Conclusion}

The main result of this work is the derivation of the relic abundance
of an irreducible population of vortons not considered so far. These
vortons are continuously created by the scaling string network at all
times during the cosmological expansion and allow us to probe new
regions of the parameter space $(G\U,1/\NNstar)$, namely energy scales
that spawn the entire spectrum from $\TeV$ scales to the Planck
scale. In particular, vortons are a viable dark matter candidate for
all possible value of $G\U$ (with, however, some quite tuned values of
$\NNstar)$. We have derived their number density distribution at all
times, which is the quantity of interest for dark matter direct
detection searches~\cite{Bonazzola:1997tk, Arina:2011si,Arina:2013jma},
and derived the relevant cosmological constraints, summarized in
figures~\ref{fig:omegatotvv} and \ref{fig:depcorr}.

Throughout this work, we have, however, assumed that all the scaling
loops are produced at the same size $\alpha t$. A more complete
analysis would take into account the fact that the loop production
function is \emph{a priori} more complicated. Due to the
proliferation of kinks on the infinite string network and the
fragmentation of large loops, we expect scaling loops to be produced
at all sizes with a power-law LPF
\begin{equation}
    \calP(\ell,t,N) = C t^{-5} \left(\frac{\ell}{t}\right)^{2\chi-3}
    \delta\negthinspace\left(N - \sqrt{\frac{\ell}{\lambda}}\right)
    \Theta(t - \tcur) \Theta(\ell -\gammac t),
\end{equation}
where $\chi$ is the so-called Polchinski-Rocha exponent and $\gammac$
is the gravitational backreaction scale.  Under this assumption, many
more small loops are produced, and one can expect some boost to the
density of vortons~\cite{Auclair:2019wcv, Auclair:2020oww}. Solving
for the vorton distribution by using a Polchinski-Rocha LPF is, however,
mathematically challenging and we have not taken this route in the
present paper.

Let us also mention that, in the present work, we have solved a
continuity equation to derive the vorton number density. This approach
is strictly equivalent to the one presented in
Ref.~\cite{Peter:2013jj}, which is based on solving a Boltzmann
equation. As a matter of fact, all the results presented have been
cross-checked using the two methods. For completeness, we give in
appendix~\ref{App1} a proof of the equivalence between the two
formalisms and how to pass from one evolution equation to the other.

Finally, concerning the influence of the initial conditions, let us
remark that in the most generic situation, one cannot exclude that the
redshift $\zini$ at which strings are formed and the redshift $\zcur$
at which the current appears are independent of the value of $G\U$ and
$\NNstar$ (or $\lambda$). Although such a situation would be difficult
to envisage for cosmic strings interpreted as topological defects, it
could be very well the case for cosmic superstrings. For instance,
$\zini$ could be very large, close to the Planck energy scales while
the warped observed value of $G\U$ can remain very low. In this case,
our assumptions of section~\ref{sec:icloop} do no longer apply and
this could change the relaxed vorton contribution. However, this would
not change the produced vorton abundance, these ones being generated
by the network at all subsequent times. A complete model-independent
treatment would require to consider a four-dimensional parameter space
made of ($G\U,\NNstar,\zini,\zcur)$, which could be explored using
Monte-Carlo-Markov-Chain methods, but we leave such a study for a future
work.

\section*{Acknowledgements}

P.~A. thanks the organizers of the Paris Primordial Cosmology Meetings
where this project germinated. P.~P. wishes to thank Churchill College,
Cambridge, where he was partially supported by a fellowship funded by
the Higher Education, Research and Innovation Department of the French
Embassy to the United-Kingdom during this research. The work of C.~R.
is supported by the ``Fonds de la Recherche Scientifique - FNRS''
under Grant $\mathrm{N^{\circ}T}.0198.19$ as well as by the
Wallonia-Brussels Federation Grant ARC $\mathrm{N^{\circ}}19/24-103$. 

\appendix

\section{Connection between the Boltzmann and continuity equations} \label{App1}

We clarify in this Appendix the equivalence between equation~(2.7) of
Ref.~\cite{Peter:2013jj} and our equation~\eqref{eq:main} to show that the
difference merely comes from the use of either lagrangian or eulerian
coordinates. In Ref.~\cite{Peter:2013jj}, one has $\ell =
\ell(\ellini,t)$, i.e., one follows the evolution of a given loop size
$\ell$ that begun with an initial value $\ellini$; somehow, the
relevant variable is $\ellini$, and the flow is lagrangian. In the
present work, the size of the loop $\ell$ is just what it is at the
time one is concerned with, with no mention of the individual loop;
this is the eulerian version.

Going from the eulerian set $\{ \ell, t\}$ to the lagrangian one $\{
\ellini, t\}$ means that for any quantity $X(\ell,t) =
X[\ell(\ellini,t)]$, one has
$$
\ud X = \left( \frac{\partial X}{\partial t}\right)_\ell \ud t + \left( \frac{\partial X}{\partial \ell}\right)_t \ud \ell = \left( \frac{\partial X}{\partial t}\right)_{\ellini} \ud t + \left( \frac{\partial X}{\partial \ellini}\right)_t \ud \ellini,
$$
with the subscript on the brackets for the partial derivatives indicating the quantity left constant for the evaluation of the derivative.
Similarly expanding the differential $\ud \ell$ and identifying the partial derivatives, one finds
\begin{equation}
    \left( \frac{\partial X}{\partial t}\right)_{\ellini} = \left(
    \frac{\partial X}{\partial t}\right)_\ell + \left( \frac{\partial
      X}{\partial \ell}\right)_t \left( \frac{\partial \ell}{\partial
      t}\right)_{\ellini} \ \ \ \hbox{and}\ \ \ \left( \frac{\partial
      X}{\partial \ellini}\right)_t = \left( \frac{\partial
      X}{\partial \ell}\right)_t \left( \frac{\partial \ell}{\partial
      \ellini}\right)_t.
    \label{partialX}
\end{equation}
One also notes that
\begin{equation}
j \equiv \frac{\ud \ell}{\ud t} =  \left( \frac{\partial \ell}{\partial \ellini}\right)_t \frac{\ud \ellini}{\ud t} + \left( \frac{\partial \ell}{\partial t}\right)_{\ellini} = \left( \frac{\partial \ell}{\partial t}\right)_{\ellini},
\label{jpartial}
\end{equation}
the final step being a consequence of the fact that in lagrangian
coordinates, $\ellini$ does not depend on time.  Combining
\eqref{jpartial} and \eqref{partialX}, one immediately gets that
\begin{equation}
    \left( \frac{\partial X}{\partial t}\right)_\ell + j \left( \frac{\partial X}{\partial \ell}\right)_t = \left( \frac{\partial X}{\partial t}\right)_{\ellini}.
\end{equation}
We are now in position to compare equation~(2.7) of
Ref.~\cite{Peter:2013jj} and our equation~\eqref{eq:main}. The former
indeed reads
\begin{equation}
\frac{\partial}{\partial t} \left(a^3 F \Jpsd \right) + j \frac{\partial}{\partial \ell} \left(a^3 F \Jpsd \right) = a^3 \mathcal{P} \Jpsd,
    \label{2.7}
\end{equation}
where $\Jpsd = \partial \ell/\partial\ellini$ accounts for phase space
distortion and we have set $F \equiv \Nn(\ell,t,N)$ for convenience.
Expanding the partial derivatives of \eqref{2.7} and simplifying by
$\Jpsd $ (assumed non vanishing), one gets
\begin{equation}
\frac{\partial}{\partial t} \left(a^3 F \right) + j \frac{\partial}{\partial \ell} \left(a^3 F \right) + \frac{a^3 f}{\Jpsd} \left[ \left( \frac{\partial \Jpsd}{\partial t} \right)_\ell + j \left( \frac{\partial \Jpsd}{\partial \ell} \right)_t \right] = a^3 \mathcal{P},
    \label{temp1}
\end{equation}
the term in square brackets being, by virtue of \eqref{partialX} and \eqref{jpartial},
simply $\left( \partial \Jpsd / \partial t \right)_{\ellini}$.  Given
the definition of $\Jpsd$ and swapping partial derivatives, it turns
out that
$$
\left( \frac{\partial \Jpsd}{\partial t} \right)_{\ellini} = \left( \frac{\partial j}{\partial \ell} \right)_t \frac{\partial \ell}{\partial\ellini}
= \Jpsd \left( \frac{\partial j}{\partial \ell} \right)_t,
$$
so that equation~\eqref{temp1} now becomes
\begin{equation}
\frac{\partial}{\partial t} \left(a^3 F \right) + j \frac{\partial}{\partial \ell} \left(a^3 F \right) + \left( a^3 F \right)\frac{\partial j}{\partial \ell}  = a^3 \mathcal{P},
\end{equation}
which is, as announced, equation~\eqref{eq:main} after grouping the
$\ell-$derivative terms and expliciting $j$ as in equation~\eqref{eq:ldot}.

To conclude this appendix, we give in Table~\ref{diffparams}, a
dictionary between the different notation used in
Refs.~\cite{Brandenberger:1996zp, Peter:2013jj} and the present work.
\begin{table}[ht]
    \centering
    \begin{tabular}{c|c}
        Present work & Ref.~\cite{Peter:2013jj} \\
        \hline
        $ \U $    & $U$ \\
        $ \ell_0$  & $\ellv$ \\
        $ \calR $ & $ N_*$ \\
        $ \gammad$ &  $\gamma_\ud$ \\
        $\sigma \to 0$  & $\gammav \to 0$\\
        $ \ell_\star = \lambda^3 \mu $ & $\ellp$ \\        
        $ \ellT(t) $ & $\ellT(t) $
    \end{tabular}
    \caption{Dictionary of notation between the present work and
      Refs.~\cite{Brandenberger:1996zp, Peter:2013jj}.}
    \label{diffparams}
\end{table}

\section{Distribution of proto-vortons and doomed loops}
\label{sec:others}

In this Appendix, we give the distributions of proto-vortons and
doomed loops, both of which contribute to the stochastic gravitational
wave background.  Proto-vortons and doomed loops decay through
gravitational wave radiation and their collapse is not prevented by
the current: indeed for both, $\calJ = 1$ (in the limit $\sigma
\rightarrow 0$). Hence for these distributions, and without loss of
generality, we set $\calJ = 1$ in this Appendix.

\subsection{Doomed loops}

Doomed loops are the loops which do not have enough current to prevent their final collapse, hence $N<\calR$.
From equations~\eqref{eq:doomdef}, \eqref{eq:sol1} and \eqref{eq:cond}, the relaxed doomed loop distribution, that is to say the doomed loops which are produced from the initial conditions at condensation, reads
\begin{equation}
    \left. \N\right|_\mathrm{doom, rel}
    = \left[\frac{a(\tcur)}{a(t)}\right]^3 \int \ud N \N\left(\lcur,\tcur\right) \delta\!\left(N - \sqrt{\frac{\lcur}{\lambda}}\right) \Theta(\calR - N),
\end{equation}
in which $\lcur$, the size of the loop during condensation at $\tcur$, is given by
\begin{equation}
    \lcur(\ell, t) = \Gamma G\mu (t - \tcur) + \ell.
\end{equation}
Integrating over the charge $N$ and replacing $\lcur$, one obtains the number density of doomed loops in relaxation
\begin{equation}
    \left.\N\right|_\mathrm{doom, rel}
    = \left[\frac{a(\tcur)}{a(t)}\right]^3 \N\left[\Gamma G\mu (t - \tcur) + \ell,\tcur\right] \Theta\!\left[\calR - \sqrt{\frac{\Gamma G\mu (t - \tcur) + \ell}{\lambda}}\right].
\end{equation}

Concerning the doomed loops produced after condensation, from equations~\eqref{eq:doomdef} and \eqref{eq:d2NdlddN}, their number density is given by
\begin{equation}
    \left.\N\right|_\mathrm{doom, prod}
    = C \int \ud N \Theta(\calR - N) \left[\frac{a(\tstar)}{a(t)}\right]^3 \tstar^{-4} \frac{1}{\alpha + \Gamma G\mu} \delta\!\left(N - \sqrt{\frac{\alpha \tstar}{\lambda}}\right) \Theta(\tstar - \tcur),
\end{equation}
in which $\tstar$ is the loop formation time.
Assuming, as we have done throughout this paper, that loops are produced at a given size $\ell = \alpha t$ at time $t$, the formation time satisfies
\begin{equation}
    \tstar(\ell, t) = \frac{\ell + \Gamma G\mu t}{\alpha + \Gamma G\mu}.
\end{equation}
Finally, integrating over the charge $N$ and replacing the formation time by the above equation, one obtains the number density of doomed loops produced after condensation:
\begin{equation}
\begin{aligned}
  \left.\N\right|_\mathrm{doom, prod}
    & = C \left[\frac{a\left(\frac{\ell + \Gamma G\mu t }{\alpha +
        \Gamma G\mu}\right)}{a(t)}\right]^3 \frac{(\alpha + \Gamma
    G\mu)^3}{(\ell + \Gamma G\mu t)^4} \\ & \times
  \Theta\!\left(\frac{\ell + \Gamma G\mu t }{\alpha + \Gamma G\mu} - \tcur\right) \Theta\!\left[\calR - \sqrt{\frac{\alpha (\ell + \Gamma G\mu t)}{\lambda(\alpha + \Gamma G\mu)}} \right].
\end{aligned}
\end{equation}

\subsection{Proto-vortons}

Proto-vortons are loops which will eventually become vortons after a
certain time, but which are still large enough to behave like
Nambu-Goto strings.  From equation~\eqref{eq:protodef},
\eqref{eq:sol1} and \eqref{eq:cond}, the distribution of ``relaxed
proto-vortons'' is given by
\begin{equation}
    \left.\N\right|_\mathrm{proto, relax} 
    = \int \ud N \Theta(N - \calR) \left[\frac{a(\tcur)}{a(t)}\right]^3 \N\left(\lcur,\tcur\right) \delta\!\left(N - \sqrt{\frac{\lcur}{\lambda}}\right) \Theta\left[\ell - \ell_0(N)\right],
    \label{eq:early}
\end{equation}
where $\ell_0(N)=N/\sqrt{\mu}$ and, again,
the size of the loop at formation is given by
\begin{equation}
    \lcur(\ell, t) = \Gamma G\mu (t - \tcur) + \ell.
\end{equation}
On carrying out the integral over the charge $N$ in equation \eqref{eq:early}, the number density of proto-vortons produced at condensation is
\begin{align}
    \left.\N\right|_\mathrm{proto, relax}
    &= \left[\frac{a(\tcur)}{a(t)}\right]^3  \N\left[\Gamma G\mu (t - \tcur) + \ell,\tcur\right] \nonumber \\ 
    & \times \Theta\!\left[\ell - \sqrt{\frac{\Gamma G\mu (t - \tcur) + \ell}{\lambda \mu}}\right] \Theta\!\left[\sqrt{\frac{\Gamma G\mu (t - \tcur) + \ell}{\lambda}} - \calR\right].
\end{align}

Proto-vortons can also be produced after condensation, in which case their distribution is obtained from equations~\eqref{eq:protodef} and \eqref{eq:d2NdlddN}
\begin{equation}
\begin{aligned}
  \left.\NN\right|_\mathrm{proto, prod} 
    & = C\int \ud N  \left[\frac{a(\tstar)}{a(t)}\right]^3 \tstar^{-4} \dfrac{\Theta(N - \calR)}{\alpha + \Gamma G\mu } \delta\!\left(N - \sqrt{\frac{\alpha \tstar}{\lambda}}\right) \Theta(\tstar - \tcur) \Theta\left[\ell - \ell_0(N)\right].
\end{aligned}
\end{equation}
Similarly, the loop formation time $\tstar$ is given by 
\begin{equation}
    \tstar(\ell, t) = \frac{\ell + \Gamma G\mu t }{\alpha + \Gamma G\mu}, \qquad \ell_0(N) = \frac{N}{\sqrt{\mu}}\,.
\end{equation}
The distribution of proto-vortons produced after the condensation now reads
\begin{align}
    \left.\NN\right|_\mathrm{proto, prod}
    &= C\left[\frac{a\left(\frac{\ell + \Gamma G\mu t}{\alpha + \Gamma G\mu}\right)}{a(t)}\right]^3 \frac{(\alpha + \Gamma G\mu)^3}{(\ell + \Gamma G\mu t)^4} \nonumber \\ 
    & \times \Theta\!\left(\frac{\ell + \Gamma G\mu t}{\alpha + \Gamma G\mu} - \tcur\right) \Theta\!\left[\ell - \sqrt{\frac{\alpha (\ell + \Gamma G\mu t)}{\lambda \mu(\alpha + \Gamma G\mu)}}\right] \Theta\!\left[\sqrt{\frac{\alpha (\ell + \Gamma G\mu t)}{\lambda \mu(\alpha + \Gamma G\mu)}} - \calR\right].
\end{align}

\bibliographystyle{JHEP}
\bibliography{ref}

\providecommand{\href}[2]{#2}\begingroup\raggedright\begin{thebibliography}{10}

\bibitem{Kibble:1976sj}
T.~Kibble, \emph{{Topology of Cosmic Domains and Strings}},
  \href{https://doi.org/10.1088/0305-4470/9/8/029}{\emph{J. Phys. A} {\bfseries
  9} (1976) 1387}.

\bibitem{Lorenz:2010sm}
L.~Lorenz, C.~Ringeval and M.~Sakellariadou, \emph{{Cosmic string loop
  distribution on all length scales and at any redshift}},
  \href{https://doi.org/10.1088/1475-7516/2010/10/003}{\emph{JCAP} {\bfseries
  10} (2010) 003} [\href{https://arxiv.org/abs/1006.0931}{{\ttfamily
  1006.0931}}].

\bibitem{Damour:2001bk}
T.~Damour and A.~Vilenkin, \emph{{Gravitational wave bursts from cusps and
  kinks on cosmic strings}},
  \href{https://doi.org/10.1103/PhysRevD.64.064008}{\emph{Phys. Rev. D}
  {\bfseries 64} (2001) 064008}
  [\href{https://arxiv.org/abs/gr-qc/0104026}{{\ttfamily gr-qc/0104026}}].

\bibitem{Blanco-Pillado:2017oxo}
J.J.~Blanco-Pillado and K.D.~Olum, \emph{{Stochastic gravitational wave
  background from smoothed cosmic string loops}},
  \href{https://doi.org/10.1103/PhysRevD.96.104046}{\emph{Phys. Rev. D}
  {\bfseries 96} (2017) 104046}
  [\href{https://arxiv.org/abs/1709.02693}{{\ttfamily 1709.02693}}].

\bibitem{Ringeval:2017eww}
C.~Ringeval and T.~Suyama, \emph{{Stochastic gravitational waves from cosmic
  string loops in scaling}},
  \href{https://doi.org/10.1088/1475-7516/2017/12/027}{\emph{JCAP} {\bfseries
  12} (2017) 027} [\href{https://arxiv.org/abs/1709.03845}{{\ttfamily
  1709.03845}}].

\bibitem{Witten:1985fp}
E.~Witten, \emph{{Cosmic Superstrings}},
  \href{https://doi.org/10.1016/0370-2693(85)90540-4}{\emph{Phys. Lett. B}
  {\bfseries 153} (1985) 243}.

\bibitem{Lazarides:1986di}
G.~Lazarides, C.~Panagiotakopoulos and Q.~Shafi, \emph{{Superheavy
  Superconducting Cosmic Strings From Superstring Models}},
  \href{https://doi.org/10.1016/0370-2693(87)90966-X}{\emph{Phys. Lett. B}
  {\bfseries 183} (1987) 289}.

\bibitem{Carter:1999pq}
B.~Carter, \emph{{Dilatonic formulation for conducting cosmic string models}},
  \href{https://doi.org/10.1002/(SICI)1521-3889(200005)9:3/5<247::AID-ANDP247>3.0.CO;2-5}{\emph{Annalen
  Phys.} {\bfseries 9} (2000) 247}
  [\href{https://arxiv.org/abs/hep-th/0002162}{{\ttfamily hep-th/0002162}}].

\bibitem{Carter:2000wv}
B.~Carter, \emph{{Essentials of classical brane dynamics}},
  \href{https://doi.org/10.1023/A:1012934901706}{\emph{Int. J. Theor. Phys.}
  {\bfseries 40} (2001) 2099}
  [\href{https://arxiv.org/abs/gr-qc/0012036}{{\ttfamily gr-qc/0012036}}].

\bibitem{Kibble:1981gv}
T.~Kibble, \emph{{Phase Transitions in the Early Universe}}, {\emph{Acta Phys.
  Polon. B} {\bfseries 13} (1982) 723}.

\bibitem{Brandenberger:1996zp}
R.H.~Brandenberger, B.~Carter, A.-C.~Davis and M.~Trodden, \emph{{Cosmic
  vortons and particle physics constraints}},
  \href{https://doi.org/10.1103/PhysRevD.54.6059}{\emph{Phys. Rev. D}
  {\bfseries 54} (1996) 6059}
  [\href{https://arxiv.org/abs/hep-ph/9605382}{{\ttfamily hep-ph/9605382}}].

\bibitem{Jones:2002cv}
N.T.~Jones, H.~Stoica and S.~Tye, \emph{{Brane interaction as the origin of
  inflation}}, \href{https://doi.org/10.1088/1126-6708/2002/07/051}{\emph{JHEP}
  {\bfseries 07} (2002) 051}
  [\href{https://arxiv.org/abs/hep-th/0203163}{{\ttfamily hep-th/0203163}}].

\bibitem{Sarangi:2002yt}
S.~Sarangi and S.~Tye, \emph{{Cosmic string production towards the end of brane
  inflation}}, \href{https://doi.org/10.1016/S0370-2693(02)01824-5}{\emph{Phys.
  Lett. B} {\bfseries 536} (2002) 185}
  [\href{https://arxiv.org/abs/hep-th/0204074}{{\ttfamily hep-th/0204074}}].

\bibitem{Urrestilla:2007yw}
J.~Urrestilla and A.~Vilenkin, \emph{{Evolution of cosmic superstring networks:
  A Numerical simulation}},
  \href{https://doi.org/10.1088/1126-6708/2008/02/037}{\emph{JHEP} {\bfseries
  02} (2008) 037} [\href{https://arxiv.org/abs/0712.1146}{{\ttfamily
  0712.1146}}].

\bibitem{Polchinski:2004yav}
J.~Polchinski, \emph{{Cosmic superstrings revisited}},
  \href{https://doi.org/10.1142/S0217751X05026686}{\emph{AIP Conf. Proc.}
  {\bfseries 743} (2004) 331}
  [\href{https://arxiv.org/abs/hep-th/0410082}{{\ttfamily hep-th/0410082}}].

\bibitem{Carter:1990bb}
B.~Carter, \emph{{Covariant Mechanics of Simple and Conducting Strings and
  Membranes}},  in \emph{{The Formation and evolution of cosmic strings.
  Proceedings, Workshop, Cambridge, UK, July 3-7, 1989}}, G.~Gibbons,
  S.~Hawking and T.~Vachaspati, eds., pp.~143--178, 1990.

\bibitem{Carter:1994hn}
B.~Carter and P.~Peter, \emph{{Supersonic string models for Witten vortices}},
  \href{https://doi.org/10.1103/PhysRevD.52.1744}{\emph{Phys.Rev.} {\bfseries
  D52} (1995) 1744} [\href{https://arxiv.org/abs/hep-ph/9411425}{{\ttfamily
  hep-ph/9411425}}].

\bibitem{Davis:1988ij}
R.~Davis and E.~Shellard, \emph{{Cosmic vortons}},
  \href{https://doi.org/10.1016/0550-3213(89)90594-4}{\emph{Nucl.Phys.}
  {\bfseries B323} (1989) 209}.

\bibitem{Carter:1990gz}
B.~Carter, \emph{{Cosmic rings as a chump dark matter candidate?}},  in
  \emph{{10th Moriond Astrophysics Meeting: The Early Phases of the Universe}},
  pp.~213--221, 1990.

\bibitem{Martins:1998gb}
C.~Martins and E.~Shellard, \emph{{Vorton formation}},
  \href{https://doi.org/10.1103/PhysRevD.57.7155}{\emph{Phys. Rev. D}
  {\bfseries 57} (1998) 7155}
  [\href{https://arxiv.org/abs/hep-ph/9804378}{{\ttfamily hep-ph/9804378}}].

\bibitem{Martins:1998th}
C.~Martins and E.~Shellard, \emph{{Limits on cosmic chiral vortons}},
  \href{https://doi.org/10.1016/S0370-2693(98)01466-X}{\emph{Phys. Lett. B}
  {\bfseries 445} (1998) 43}
  [\href{https://arxiv.org/abs/hep-ph/9806480}{{\ttfamily hep-ph/9806480}}].

\bibitem{Carter:1999wy}
B.~Carter, \emph{{Old and new processes of vorton formation}}, {\emph{Lect.
  Notes Phys.} {\bfseries 541} (2000) 71}
  [\href{https://arxiv.org/abs/hep-ph/9909513}{{\ttfamily hep-ph/9909513}}].

\bibitem{Davis:2000cx}
A.C.~Davis, T.W.B.~Kibble, M.~Pickles and D.A.~Steer, \emph{{Dynamics and
  properties of chiral cosmic strings in Minkowski space}},
  \href{https://doi.org/10.1103/PhysRevD.62.083516}{\emph{Phys. Rev. D}
  {\bfseries 62} (2000) 083516}
  [\href{https://arxiv.org/abs/astro-ph/0005514}{{\ttfamily
  astro-ph/0005514}}].

\bibitem{Steer:2000jn}
D.A.~Steer, \emph{{Selfintersections and gravitational properties of chiral
  cosmic strings in Minkowski space}},
  \href{https://doi.org/10.1103/PhysRevD.63.083517}{\emph{Phys. Rev. D}
  {\bfseries 63} (2001) 083517}
  [\href{https://arxiv.org/abs/astro-ph/0011233}{{\ttfamily
  astro-ph/0011233}}].

\bibitem{Lemperiere:2003yt}
Y.~Lemperiere and E.~Shellard, \emph{{Vorton existence and stability}},
  \href{https://doi.org/10.1103/PhysRevLett.91.141601}{\emph{Phys. Rev. Lett.}
  {\bfseries 91} (2003) 141601}
  [\href{https://arxiv.org/abs/hep-ph/0305156}{{\ttfamily hep-ph/0305156}}].

\bibitem{Battye:2008mm}
R.A.~Battye and P.M.~Sutcliffe, \emph{{Vorton construction and dynamics}},
  \href{https://doi.org/10.1016/j.nuclphysb.2009.01.021}{\emph{Nucl. Phys. B}
  {\bfseries 814} (2009) 180}
  [\href{https://arxiv.org/abs/0812.3239}{{\ttfamily 0812.3239}}].

\bibitem{Garaud:2013iba}
J.~Garaud, E.~Radu and M.S.~Volkov, \emph{{Stable Cosmic Vortons}},
  \href{https://doi.org/10.1103/PhysRevLett.111.171602}{\emph{Phys. Rev. Lett.}
  {\bfseries 111} (2013) 171602}
  [\href{https://arxiv.org/abs/1303.3044}{{\ttfamily 1303.3044}}].

\bibitem{Peter:2013jj}
P.~Peter and C.~Ringeval, \emph{{A Boltzmann treatment for the vorton excess
  problem}}, \href{https://doi.org/10.1088/1475-7516/2013/05/005}{\emph{JCAP}
  {\bfseries 05} (2013) 005} [\href{https://arxiv.org/abs/1302.0953}{{\ttfamily
  1302.0953}}].

\bibitem{Ringeval:2005kr}
C.~Ringeval, M.~Sakellariadou and F.~Bouchet, \emph{{Cosmological evolution of
  cosmic string loops}}, {\emph{JCAP} {\bfseries 0702} (2007) 023}
  [\href{https://arxiv.org/abs/astro-ph/0511646}{{\ttfamily
  astro-ph/0511646}}].

\bibitem{Bonazzola:1997tk}
S.~Bonazzola and P.~Peter, \emph{{Can high-energy cosmic rays be vortons?}},
  \href{https://doi.org/10.1016/S0927-6505(97)00015-7}{\emph{Astropart. Phys.}
  {\bfseries 7} (1997) 161}
  [\href{https://arxiv.org/abs/hep-ph/9701246}{{\ttfamily hep-ph/9701246}}].

\bibitem{Aghanim:2018eyx}
{\scshape Planck} collaboration, \emph{{Planck 2018 results. VI. Cosmological
  parameters}},
  \href{https://doi.org/10.1051/0004-6361/201833910}{\emph{Astron. Astrophys.}
  {\bfseries 641} (2020) A6}
  [\href{https://arxiv.org/abs/1807.06209}{{\ttfamily 1807.06209}}].

\bibitem{Auclair:2019jip}
P.~Auclair, D.A.~Steer and T.~Vachaspati, \emph{{Particle emission and
  gravitational radiation from cosmic strings: observational constraints}},
  \href{https://doi.org/10.1103/PhysRevD.101.083511}{\emph{Phys. Rev. D}
  {\bfseries 101} (2020) 083511}
  [\href{https://arxiv.org/abs/1911.12066}{{\ttfamily 1911.12066}}].

\bibitem{Carter:1989dp}
B.~Carter, \emph{{Duality relation between charged elastic strings and
  superconducting cosmic strings}},
  \href{https://doi.org/10.1016/0370-2693(89)91051-4}{\emph{Phys. Lett.}
  {\bfseries B224} (1989) 61}.

\bibitem{Carter:1992ny}
B.~Carter, \emph{{Basic brane theory}},
  \href{https://doi.org/10.1088/0264-9381/9/S/002}{\emph{Class. Quant. Grav.}
  {\bfseries 9} (1992) S19}.

\bibitem{Babul:1987me}
A.~Babul, T.~Piran and D.N.~Spergel, \emph{{Bosonic superconducting cosmic
  strings. 1. Classical field theory solutions}},
  \href{https://doi.org/10.1016/0370-2693(88)90476-5}{\emph{Phys. Lett. B}
  {\bfseries 202} (1988) 307}.

\bibitem{Peter:1992dw}
P.~Peter, \emph{{Superconducting cosmic string: Equation of state for space -
  like and time - like current in the neutral limit}},
  \href{https://doi.org/10.1103/PhysRevD.45.1091}{\emph{Phys. Rev.} {\bfseries
  D45} (1992) 1091}.

\bibitem{Carter:1999hx}
B.~Carter and P.~Peter, \emph{{Dynamics and integrability property of the
  chiral string model}},
  \href{https://doi.org/10.1016/S0370-2693(99)01070-9}{\emph{Phys. Lett.}
  {\bfseries B466} (1999) 41}
  [\href{https://arxiv.org/abs/hep-th/9905025}{{\ttfamily hep-th/9905025}}].

\bibitem{Ringeval:2000kz}
C.~Ringeval, \emph{{Equation of state of cosmic strings with fermionic
  current-carriers}},
  \href{https://doi.org/10.1103/PhysRevD.63.063508}{\emph{Phys. Rev.}
  {\bfseries D63} (2001) 063508}
  [\href{https://arxiv.org/abs/hep-ph/0007015}{{\ttfamily hep-ph/0007015}}].

\bibitem{Ringeval:2001xd}
C.~Ringeval, \emph{{Fermionic massive modes along cosmic strings}},
  \href{https://doi.org/10.1103/PhysRevD.64.123505}{\emph{Phys. Rev.}
  {\bfseries D64} (2001) 123505}
  [\href{https://arxiv.org/abs/hep-ph/0106179}{{\ttfamily hep-ph/0106179}}].

\bibitem{Peter:1992nz}
P.~Peter, \emph{{Equation of state of cosmic strings in the presence of charged
  particles}}, \href{https://doi.org/10.1088/0264-9381/9/S/013}{\emph{Class.
  Quant. Grav.} {\bfseries 9} (1992) S197}.

\bibitem{Peter:1992ta}
P.~Peter, \emph{{Influence of the electric coupling strength in current
  carrying cosmic strings}},
  \href{https://doi.org/10.1103/PhysRevD.46.3335}{\emph{Phys. Rev. D}
  {\bfseries 46} (1992) 3335}.

\bibitem{Carter:1990sm}
B.~Carter, \emph{{Mechanics of cosmic rings}},
  \href{https://doi.org/10.1016/0370-2693(90)91714-M}{\emph{Phys. Lett. B}
  {\bfseries 238} (1990) 166}
  [\href{https://arxiv.org/abs/hep-th/0703023}{{\ttfamily hep-th/0703023}}].

\bibitem{Davis:1997bs}
S.C.~Davis, A.-C.~Davis and M.~Trodden, \emph{{N=1 supersymmetric cosmic
  strings}}, \href{https://doi.org/10.1016/S0370-2693(97)00642-4}{\emph{Phys.
  Lett. B} {\bfseries 405} (1997) 257}
  [\href{https://arxiv.org/abs/hep-ph/9702360}{{\ttfamily hep-ph/9702360}}].

\bibitem{Carter:2003fb}
B.~Carter and D.A.~Steer, \emph{{Symplectic structure for elastic and chiral
  conducting cosmic string models}},
  \href{https://doi.org/10.1103/PhysRevD.69.125002}{\emph{Phys. Rev.}
  {\bfseries D69} (2004) 125002}
  [\href{https://arxiv.org/abs/hep-th/0307161}{{\ttfamily hep-th/0307161}}].

\bibitem{PhysRevD.45.1898}
B.~Allen and E.P.S.~Shellard, \emph{Gravitational radiation from cosmic
  strings}, \href{https://doi.org/10.1103/PhysRevD.45.1898}{\emph{Phys. Rev. D}
  {\bfseries 45} (1992) 1898}.

\bibitem{Copeland:1998na}
E.J.~Copeland, T.~Kibble and D.A.~Steer, \emph{{The Evolution of a network of
  cosmic string loops}},
  \href{https://doi.org/10.1103/PhysRevD.58.043508}{\emph{Phys. Rev. D}
  {\bfseries 58} (1998) 043508}
  [\href{https://arxiv.org/abs/hep-ph/9803414}{{\ttfamily hep-ph/9803414}}].

\bibitem{Rocha:2007ni}
J.V.~Rocha, \emph{{Scaling solution for small cosmic string loops}},
  \href{https://doi.org/10.1103/PhysRevLett.100.071601}{\emph{Phys. Rev. Lett.}
  {\bfseries 100} (2008) 071601}
  [\href{https://arxiv.org/abs/0709.3284}{{\ttfamily 0709.3284}}].

\bibitem{Auclair:2019zoz}
P.~Auclair, C.~Ringeval, M.~Sakellariadou and D.~Steer, \emph{{Cosmic string
  loop production functions}},
  \href{https://doi.org/10.1088/1475-7516/2019/06/015}{\emph{JCAP} {\bfseries
  06} (2019) 015} [\href{https://arxiv.org/abs/1903.06685}{{\ttfamily
  1903.06685}}].

\bibitem{Polchinski:2006ee}
J.~Polchinski and J.V.~Rocha, \emph{Analytic study of small scale structure on
  cosmic strings}, {\emph{Phys. Rev.} {\bfseries D74} (2006) 083504}
  [\href{https://arxiv.org/abs/hep-ph/0606205}{{\ttfamily hep-ph/0606205}}].

\bibitem{Dubath:2007mf}
F.~Dubath, J.~Polchinski and J.V.~Rocha, \emph{{Cosmic String Loops, Large and
  Small}}, \href{https://doi.org/10.1103/PhysRevD.77.123528}{\emph{Phys. Rev.}
  {\bfseries D77} (2008) 123528}
  [\href{https://arxiv.org/abs/0711.0994}{{\ttfamily 0711.0994}}].

\bibitem{Auclair:2020oww}
P.G.~Auclair, \emph{{Impact of the small-scale structure on the Stochastic
  Background of Gravitational Waves from cosmic strings}},
  \href{https://doi.org/10.1088/1475-7516/2020/11/050}{\emph{JCAP} {\bfseries
  11} (2020) 050} [\href{https://arxiv.org/abs/2009.00334}{{\ttfamily
  2009.00334}}].

\bibitem{Siemens:2002dj}
X.~Siemens, K.D.~Olum and A.~Vilenkin, \emph{{On the size of the smallest
  scales in cosmic string networks}},
  \href{https://doi.org/10.1103/PhysRevD.66.043501}{\emph{Phys. Rev.}
  {\bfseries D66} (2002) 043501}
  [\href{https://arxiv.org/abs/gr-qc/0203006}{{\ttfamily gr-qc/0203006}}].

\bibitem{Polchinski:2007rg}
J.~Polchinski and J.V.~Rocha, \emph{{Cosmic string structure at the
  gravitational radiation scale}},
  \href{https://doi.org/10.1103/PhysRevD.75.123503}{\emph{Phys. Rev.}
  {\bfseries D75} (2007) 123503}
  [\href{https://arxiv.org/abs/gr-qc/0702055}{{\ttfamily gr-qc/0702055}}].

\bibitem{Vachaspati:1984}
T.~{Vachaspati} and A.~{Vilenkin}, \emph{{Formation and evolution of cosmic
  strings}}, {\emph{Phys. Rev.} {\bfseries D30} (1984) 2036}.

\bibitem{Hindmarsh:2005ix}
M.~Hindmarsh and O.~Philipsen, \emph{{WIMP dark matter and the QCD equation of
  state}}, \href{https://doi.org/10.1103/PhysRevD.71.087302}{\emph{Phys. Rev.}
  {\bfseries D71} (2005) 087302}
  [\href{https://arxiv.org/abs/hep-ph/0501232}{{\ttfamily hep-ph/0501232}}].

\bibitem{2003PhRvD..68j3514J}
R.~{Jeannerot}, J.~{Rocher} and M.~{Sakellariadou}, \emph{{How generic is
  cosmic string formation in supersymmetric grand unified theories}},
  \href{https://doi.org/10.1103/PhysRevD.68.103514}{\emph{Phys. Rev.}
  {\bfseries D68} (2003) 103514}
  [\href{https://arxiv.org/abs/hep-ph/0308134}{{\ttfamily hep-ph/0308134}}].

\bibitem{2005JCAP...03..004R}
J.~{Rocher} and M.~{Sakellariadou}, \emph{{Constraints on supersymmetric grand
  unified theories from cosmology}},
  \href{https://doi.org/10.1088/1475-7516/2005/03/004}{\emph{JCAP} {\bfseries
  2005} (2005) 004} [\href{https://arxiv.org/abs/hep-ph/0406120}{{\ttfamily
  hep-ph/0406120}}].

\bibitem{Rajantie:2001ps}
A.~Rajantie, \emph{{Formation of topological defects in gauge field theories}},
  \href{https://doi.org/10.1142/S0217751X02005426}{\emph{Int. J. Mod. Phys. A}
  {\bfseries 17} (2002) 1}
  [\href{https://arxiv.org/abs/hep-ph/0108159}{{\ttfamily hep-ph/0108159}}].

\bibitem{Rivers:2002vm}
R.~Rivers, F.~Lombardo and F.~Mazzitelli, \emph{{The Formation of classical
  defects after a slow quantum phase transition}},
  \href{https://doi.org/10.1016/S0370-2693(02)02044-0}{\emph{Phys. Lett. B}
  {\bfseries 539} (2002) 1}
  [\href{https://arxiv.org/abs/hep-ph/0205337}{{\ttfamily hep-ph/0205337}}].

\bibitem{2016JCAP...02..033R}
C.~{Ringeval}, D.~{Yamauchi}, J.~{Yokoyama} and F.R.~{Bouchet}, \emph{{Large
  scale CMB anomalies from thawing cosmic strings}},
  \href{https://doi.org/10.1088/1475-7516/2016/02/033}{\emph{JCAP} {\bfseries
  2016} (2016) 033} [\href{https://arxiv.org/abs/1510.01916}{{\ttfamily
  1510.01916}}].

\bibitem{Blanco-Pillado:2017rnf}
J.J.~Blanco-Pillado, K.D.~Olum and X.~Siemens, \emph{{New limits on cosmic
  strings from gravitational wave observation}},
  \href{https://doi.org/10.1016/j.physletb.2018.01.050}{\emph{Phys. Lett. B}
  {\bfseries 778} (2018) 392}
  [\href{https://arxiv.org/abs/1709.02434}{{\ttfamily 1709.02434}}].

\bibitem{Abbott:2017mem}
{\scshape LIGO Scientific, Virgo} collaboration, \emph{{Constraints on cosmic
  strings using data from the first Advanced LIGO observing run}},
  \href{https://doi.org/10.1103/PhysRevD.97.102002}{\emph{Phys. Rev.}
  {\bfseries D97} (2018) 102002}
  [\href{https://arxiv.org/abs/1712.01168}{{\ttfamily 1712.01168}}].

\bibitem{Ringeval:2012tk}
C.~Ringeval and F.R.~Bouchet, \emph{{All Sky CMB Map from Cosmic Strings
  Integrated Sachs-Wolfe Effect}},
  \href{https://doi.org/10.1103/PhysRevD.86.023513}{\emph{Phys.Rev.} {\bfseries
  D86} (2012) 023513} [\href{https://arxiv.org/abs/1204.5041}{{\ttfamily
  1204.5041}}].

\bibitem{Ade:2013xla}
{\scshape Planck} collaboration, \emph{{Planck 2013 results. XXV. Searches for
  cosmic strings and other topological defects}},
  \href{https://doi.org/10.1051/0004-6361/201321621}{\emph{Astron. Astrophys.}
  {\bfseries 571} (2014) A25}
  [\href{https://arxiv.org/abs/1303.5085}{{\ttfamily 1303.5085}}].

\bibitem{Lizarraga:2014xza}
J.~Lizarraga, J.~Urrestilla, D.~Daverio, M.~Hindmarsh, M.~Kunz et~al.,
  \emph{{Constraining topological defects with temperature and polarization
  anisotropies}},
  \href{https://doi.org/10.1103/PhysRevD.90.103504}{\emph{Phys.Rev.} {\bfseries
  D90} (2014) 103504} [\href{https://arxiv.org/abs/1408.4126}{{\ttfamily
  1408.4126}}].

\bibitem{Lazanu:2014eya}
A.~Lazanu and P.~Shellard, \emph{{Constraints on the Nambu-Goto cosmic string
  contribution to the CMB power spectrum in light of new temperature and
  polarisation data}},
  \href{https://doi.org/10.1088/1475-7516/2015/02/024}{\emph{JCAP} {\bfseries
  1502} (2015) 024} [\href{https://arxiv.org/abs/1410.5046}{{\ttfamily
  1410.5046}}].

\bibitem{Lazanu:2014xxa}
A.~Lazanu, E.~Shellard and M.~Landriau, \emph{{CMB power spectrum of Nambu-Goto
  cosmic strings}},
  \href{https://doi.org/10.1103/PhysRevD.91.083519}{\emph{Phys. Rev.}
  {\bfseries D91} (2015) 083519}
  [\href{https://arxiv.org/abs/1410.4860}{{\ttfamily 1410.4860}}].

\bibitem{Cai:2011bi}
Y.-F.~Cai, E.~Sabancilar and T.~Vachaspati, \emph{{Radio bursts from
  superconducting strings}},
  \href{https://doi.org/10.1103/PhysRevD.85.023530}{\emph{Phys. Rev. D}
  {\bfseries 85} (2012) 023530}
  [\href{https://arxiv.org/abs/1110.1631}{{\ttfamily 1110.1631}}].

\bibitem{Cai:2012zd}
Y.-F.~Cai, E.~Sabancilar, D.A.~Steer and T.~Vachaspati, \emph{{Radio Broadcasts
  from Superconducting Strings}},
  \href{https://doi.org/10.1103/PhysRevD.86.043521}{\emph{Phys. Rev. D}
  {\bfseries 86} (2012) 043521}
  [\href{https://arxiv.org/abs/1205.3170}{{\ttfamily 1205.3170}}].

\bibitem{Arina:2011si}
C.~Arina, J.~Hamann and Y.Y.~Wong, \emph{{A Bayesian view of the current status
  of dark matter direct searches}},
  \href{https://doi.org/10.1088/1475-7516/2011/09/022}{\emph{JCAP} {\bfseries
  09} (2011) 022} [\href{https://arxiv.org/abs/1105.5121}{{\ttfamily
  1105.5121}}].

\bibitem{Arina:2013jma}
C.~Arina, \emph{{Bayesian analysis of multiple direct detection experiments}},
  \href{https://doi.org/10.1016/j.dark.2014.03.003}{\emph{Phys. Dark Univ.}
  {\bfseries 5-6} (2014) 1} [\href{https://arxiv.org/abs/1310.5718}{{\ttfamily
  1310.5718}}].

\bibitem{Auclair:2019wcv}
P.~Auclair et~al., \emph{{Probing the gravitational wave background from cosmic
  strings with LISA}},
  \href{https://doi.org/10.1088/1475-7516/2020/04/034}{\emph{JCAP} {\bfseries
  04} (2020) 034} [\href{https://arxiv.org/abs/1909.00819}{{\ttfamily
  1909.00819}}].

\end{thebibliography}\endgroup

\end{document}